\pdfoutput=1

\documentclass[twocolumn]{aastex63}
\usepackage{natbib}
\usepackage{hyperref}
\usepackage{lineno}
\usepackage{ulem}

\def\deg{\ifmmode^\circ\else$^\circ$\fi}

\shorttitle{Probing massive star formation in S284-RE}
\shortauthors{O.~R. Jadhav et al.}

\begin{document}
%

%\title{Probing the Heart of Monoceros R2: 
\title{Uncovering the hidden physical structures and protostellar activities in the Low-Metallicity S284-RE region: results from ALMA and \emph{JWST}}
\correspondingauthor{O.~R. Jadhav}
\email{Email: omkar@prl.res.in}

\author[0009-0001-2896-1896]{O.~R. Jadhav}
\affiliation{Astronomy \& Astrophysics Division, Physical Research Laboratory, Navrangpura, Ahmedabad 380009, India.}
\affiliation{Indian Institute of Technology Gandhinagar Palaj, Gandhinagar 382355, India.}

\author[0000-0001-6725-0483]{L.~K. Dewangan}
\affiliation{Astronomy \& Astrophysics Division, Physical Research Laboratory, Navrangpura, Ahmedabad 380009, India.}

\author[0000-0002-6586-936X]{Aayushi Verma}
\affiliation{Aryabhatta Research Institute of Observational Sciences, Manora Peak, Nainital 263002, India.}
\affiliation{M.J.P. Rohilkhand University, Bareilly, Uttar Pradesh-243006, India.}

\author[0000-0001-8812-8460]{N.~K.~Bhadari}
\affiliation{Astronomy \& Astrophysics Division, Physical Research Laboratory, Navrangpura, Ahmedabad 380009, India.}

\author[0000-0002-7367-9355]{A.~K. Maity}
\affiliation{Astronomy \& Astrophysics Division, Physical Research Laboratory, Navrangpura, Ahmedabad 380009, India.}
\affiliation{Indian Institute of Technology Gandhinagar Palaj, Gandhinagar 382355, India.}

\author[0000-0001-5731-3057]{Saurabh Sharma}
\affiliation{Aryabhatta Research Institute of Observational Sciences, Manora Peak, Nainital 263002, India.}

\author{Mamta}
\affiliation{Aryabhatta Research Institute of Observational Sciences, Manora Peak, Nainital 263002, India.}

% Abstract of the paper
\begin{abstract}
We present an observational study of the S284-RE region, a low-metallicity area associated with the extended S284 H\,{\sc ii} region.
A thermally supercritical filament (mass $\sim$2402 $M_{\odot}$, length $\sim$8.5 pc) is investigated using the {\it Herschel} column density map. 
The {\it Spitzer} ratio 4.5 $\mu$m/3.6 $\mu$m map traces the H$_{2}$ outflows in this filament, where previously reported young stellar objects (YSOs) are spatially distributed. 
Analysis of the YSO distribution has revealed three active star-forming clusters (YCl1, YCl2, YCl3) within the filament. 
YCl3 seems to be the most evolved, YCl2 the youngest, while YCl1 displays signs of non-thermal fragmentation. The {\it JWST} (F470N+F444W)/F356W ratio map reveals at least seven bipolar H$_{2}$ outflows, with four (olc1--olc4) in YCl1 and three (ol1--ol3) in YCl2. 
The driving sources of these outflows are identified based on outflow geometry, ALMA continuum peaks, and YSO positions. Two ALMA continuum sources, \#2 and \#3, from the $M$-$R_{\rm eff}$ plot are recognized as potential massive star formation candidates. 
The ALMA continuum source \#2 hosts at least three outflow-driving sources, whereas the ALMA continuum source \#3 contains two. The bipolar outflow olc1, driven by an embedded object within the continuum source \#2, is likely a massive protostar, as indicated by Br-$\alpha$ and PAH emissions depicted in the {\it JWST} (F405N+F444W)/F356W ratio map. The presence of H$_{2}$ knots  in the outflows olc1 and ol1 suggests episodic accretion. Overall, the study investigates a massive protostar candidate, driving the $\sim$2.7 pc 
H$_{2}$ outflow olc1 and undergoing episodic accretion.
\end{abstract}
%------------------
%
%
\keywords{
dust, extinction -- HII regions -- ISM: clouds -- ISM: individual object (Sh 2-284) -- 
stars: formation -- stars: pre--main sequence
}
%
%%%%%%%% INTRODUCTION %%%%%%%%%%%
\section{Introduction}
\label{sec:intro}
Massive OB-stars ($>$ 8 M$_{\odot}$) significantly influence their surrounding environment by releasing radiative and mechanical energy throughout their lifecycle, from formation to death. However, our understanding of the formation mechanisms of such massive stars and their feedback processes remains incomplete \citep{zinnecker07,tan14,Motte+2018,rosen20}. Over the past decade, {\it Spitzer} and {\it Herschel} observations have provided invaluable opportunities to explore key structures in star-forming regions, such as bubbles \citep[e.g.,][]{churchwell06,churchwell07} and filaments \citep[e.g.,][]{andre10,andre14,Motte+2018}. These structures play a crucial role in the processes of star formation \citep[e.g.,][]{deharveng10,Motte+2018}. 
Now, by combining data from the James Webb Space Telescope ({\it JWST}) with the Atacama Large Millimeter/submillimeter Array (ALMA), researchers have access to a powerful multi-wavelength approach \citep[e.g.,][]{dewangan24,habart24}, which can be used to investigate the dust and gas structures around embedded protostars in exceptional detail and to study distant star-forming sites. Studies have shown that low-metallicity environments are more favorable for forming massive stars due to reduced cooling rates and higher gas temperatures compared to metal-rich regions. The warmer conditions inhibit the fragmentation of gas clouds into small clumps, allowing the gas to collapse into larger structures, which can lead to the formation of more massive stars \citep[][and references therein]{larmers_2005,szecsi_2015}. As a result, such environments serve as ideal laboratories to understand the physical processes involved in massive star formation, including outflows and the accretion during the early stages of massive star formation.

This paper focuses on such a sub-region of a large H\,{\sc ii} region Sh 2-284 (or S284), characterized by a lower metallicity \citep[i.e., $Z$ $\sim$0.2 $Z_{\odot}$; see][and references therein]{kalari15}. Situated at a distance of $\sim$5.0 kpc \citep[][hereafter Puga09]{puga09}, the extended S284 H\,{\sc ii} region 
\citep[radius $\sim$13.5 pc;][]{cusano11} is located in the outer regions of the Milky Way, and hosts massive OB-stars. The site S284 is characterized by features such as filaments, pillars, and globules, which are hallmarks of star-forming regions shaped by stellar feedback  \citep{puga09,delgado10,cusano11,kalari15}. Using the {\it Spitzer} near-infrared (NIR) and mid-infrared (MIR) data, Puga09 conducted a study of young stellar objects (YSOs) in the entire site S284. They also selected several sub-regions (i.e., Dolidze 25, Cl2, Cl3, RN, RW, RS and RE; see Figure 2 in Puga09) in S284 and examined the spatial distribution of the YSOs toward these sub-regions. These authors also noted that star formation has taken place in several areas in the S284 H\,{\sc ii} region over the past few million years, and that triggered star formation is likely occurring in some of these sub-regions. 

According to \citet[][and references therein]{cusano11}, about 10 million years ago, the first generation of stars in S284 formed within a cluster. The intense UV radiation and stellar winds from the massive stars in this cluster pushed away the surrounding interstellar material, leading to the formation of stars now associated with Dolidze 25 (hereafter Dol 25). The massive stars of Dol 25 subsequently excited the S284 H\,{\sc ii} region, which further triggered the formation of additional structures, including the sub-regions. Consequently, these sub-regions merit deeper investigation with high-resolution infrared and millimeter wavelength data sets. Notably, data from both the {\it JWST} and ALMA are available only for the sub-region, S284-RE, where YSOs are aligned in a trail and no radio continuum emission is detected (see Figures 6 and 7 in Puga09). Figure~\ref{fg1}a presents the {\it Herschel} image at 160 $\mu$m overlaid with the radio continuum emission contours at 4850 MHz, revealing the extended S284 H\,{\sc ii} region and the locations of these sub-regions including S284-RE. 

This work focuses to probe the ongoing physical processes shaping star formation in this low-metallicity S284-RE region. Star-forming regions with low metallicity offer a unique opportunity to asses star formation theories across diverse environments. The S284-RE region provides valuable insights into how star formation processes differ in low-metallicity environments compared to those with solar or near-solar metallicity. Unlike solar-metallicity regions, filaments in low-metallicity environments tend to remain intact without significant fragmentation. This phenomenon supports the buildup of material in larger reservoirs, which could enhance the likelihood of massive star formation \citep{larmers_2005,szecsi_2015}. Multi-scale and multi-wavelength data sets are used to explore the underlying physical structures and protostellar activities in this site. In this connection, we employed data sets from the {\it Spitzer}, {\it Herschel}, {\it Planck}, ALMA, and {\it JWST} facilities (see Table~\ref{tab1}). The {\it Spitzer} ratio map and the {\it JWST} narrow band images are used to infer the presence of ionized emission, molecular emission, and outflows. A clustering analysis has also been performed on the previously reported YSOs toward S284-RE. Additionally, hidden physical structures at multiple scales in S284-RE are probed using the ALMA 1.38 mm continuum map and the {\it Herschel} FIR and submillimeter images. The polarimetric data from the {\it Planck} are used to deduce the orientation of the large-scale magnetic field (B-field) in the direction of S284-RE.

The structure of this paper is as follows. 
Section~\ref{sec:obser} describes the observational data sets utilized in this study. 
Our observational outcomes are presented in Section~\ref{sec:data}. 
Section~\ref{sec:disc} discusses the implications of our results for S284-RE. Finally, Section~\ref{sec:conc} summarizes the conclusions of this work.
\section{Data sets}
\label{sec:obser}
The data sets utilized in this paper are listed in Table~\ref{tab1}. The science-ready ALMA\footnote[1]{https://almascience.nao.ac.jp/aq/} 1.38 mm continuum datasets in band-6 (Project \#2021.1.01706.S; PI: Cheng, Yu) and the level-3 science ready \emph{JWST\footnote[2]{https://archive.stsci.edu/missions-and-data/jwst}} Near-Infrared Camera \citep[NIRCam;][]{2005SPIE.5904....1R, 2012SPIE.8442E..2NB} images (LW: F356W, F405N, F470N, SW:F162M, F182M, F200W; program ID \#2317; PI: Cheng, Yu; \dataset[10.17909/ea2v-0g53]{http://dx.doi.org/10.17909/ea2v-0g53}) were explored  toward S284-RE (see Table~\ref{tab1}). The positions of Class~I and Class~II YSOs, identified using the {\it Spitzer} data, were also collected from Puga09. 
\begin{table*}
 \tiny
%\scriptsize
%\small
\setlength{\tabcolsep}{0.12in}
%\small
%\scriptsize
%\setlength{\tabcolsep}{0.18in}
\centering
\caption{Different surveys employed in this paper.}
\label{tab1}
\begin{tabular}{lcccr}
\hline 
  Survey/facility  &  Wavelength/      &  Resolution        &  Reference \\   
    &  Frequency/line(s)       &   ($\arcsec$)        &   \\   
\hline
\hline 
The NRAO 4850 MHz survey                 &4850 MHz                     & $\sim$420        &\citet{condon91}\\
ALMA data    & 1.38 mm  & $\sim$0.63--0.86 & Project \#2021.1.01706.S; PI: Cheng, Yu\\
 Planck legacy archive                 &850 $\mu$m                    & $\sim$300        &\citet{planck16}\\
{\it Herschel} Infrared Galactic Plane Survey (Hi-GAL)                              &70--500 $\mu$m                     & $\sim$6--37         &\citet{molinari10}\\
Warm-{\it Spitzer} GLIMPSE360 survey       & 3.6, 4.5  $\mu$m                   & $\sim$2           &\citet{whitney11}\\
\emph{JWST} NIRCam Long Wavelength (LW) F356W, F405N, F470N imaging facility & 3.566, 4.053, 4.708 $\mu$m                   
& $\sim$0.17          &Proposal ID: 2317; PI: Cheng, Yu\\ 
\emph{JWST} NIRCam Short Wavelength (SW) F162M, F182M, F200W imaging facility  & 1.627, 1.845, 1.988 $\mu$m                   
& $\sim$0.07          &Proposal ID: 2317; PI: Cheng, Yu\\ 
\hline          
\end{tabular}
\end{table*} 
%======================================================
%
\section{Results}
\label{sec:data}
Figure~\ref{fg1}a displays the {\it Herschel} image at 160 $\mu$m overlaid with the 4850 MHz radio continuum emission contours, displaying the large-scale morphology of bubble and its edges. The target area, S284-RE is indicated by a solid box in Figure~\ref{fg1}a, and is not located at the edges of the bubble morphology. 
\subsection{{\it Spitzer} ratio map}
\label{sec:data1}
\begin{figure*}
\center
\includegraphics[width=13cm]{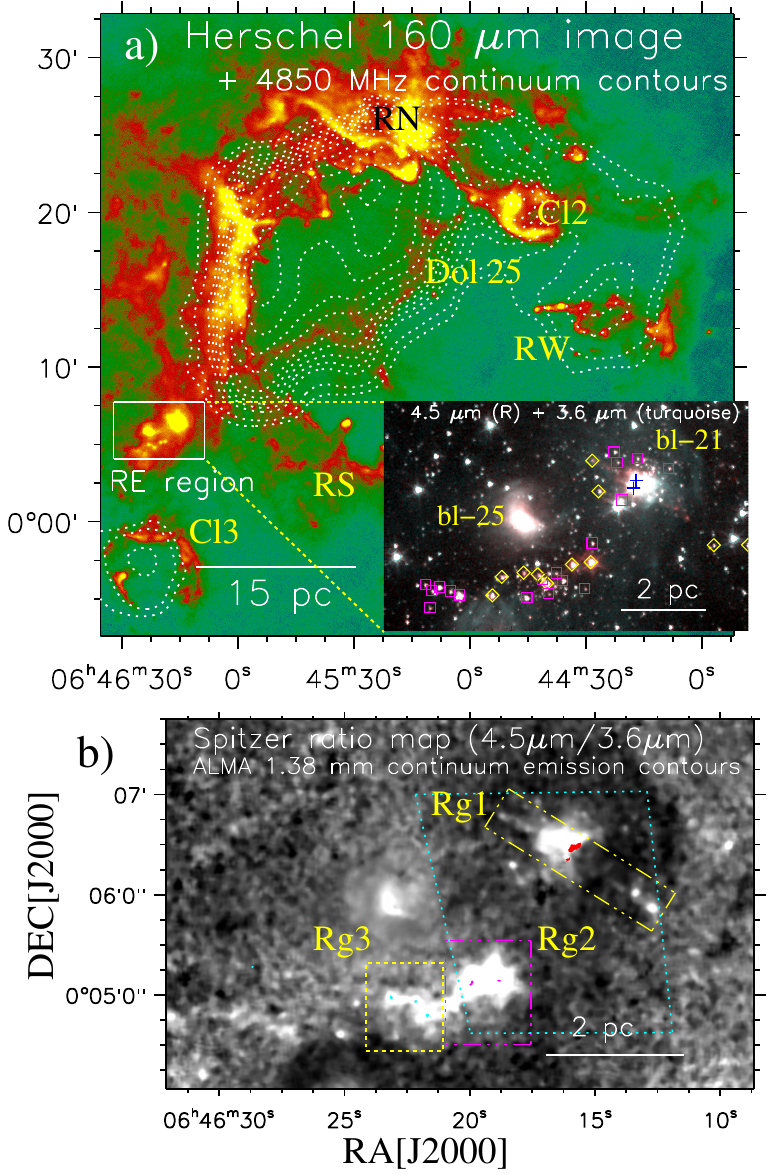}
\caption{a) Overlay of the 4850 MHz radio continuum emission contours on the {\it Herschel} 160 $\mu$m image of the large H\,{\sc ii} region S284. 
The levels of the radio continuum contours are 0.23 Jy beam$^{-1}$ $\times$ (0.15, 0.2, 0.3, 0.35, 0.4, 0.45, 0.5, 0.6, 0.7, 0.8, 0.9, 0.98). 
Several sub-regions (i.e., Dol 25, Cl2, Cl3, RN, RS, RE, and RW) studied in Puga09 are labeled. 
An inset on the bottom right displays a zoomed-in view of the selected RE sub-region (see the solid box in Figure~\ref{fg1}a). 
The inset is a two-color composite map made using the 4.5 $\mu$m (in red) and 3.6 $\mu$m (in turquoise) images, and is overlaid with the 
positions of Class~I YSOs (diamonds) and Class~II YSOs (squares) obtained from Puga09. 
Two blobs (i.e., bl-21 and bl-25) studied in Puga09 are also labeled. 
b) Overlay of the ALMA 1.38 mm continuum emission contours (in red, cyan, and magenta) on the {\it Spitzer}-GLIMPSE360 ratio map 
of 4.5 $\mu$m/3.6 $\mu$m emission toward S284-RE region. Three sub-regions Rg1, Rg2, and Rg3 are also indicated and labeled (see dashed and dot-dashed boxes). 
The dotted box (in cyan) highlights the area covered by the {\it JWST} observations. 
The level of the filled red contour is 0.68 mJy beam $^{-1}$. 
The level of the filled cyan contour is 0.40  mJy beam $^{-1}$, while the level of the filled magenta contour is 0.46 mJy beam $^{-1}$. 
Scale bars corresponding to 2 pc and 15 pc are derived at a distance of 5 kpc.} 
\label{fg1}
\end{figure*}

The inset in Figure~\ref{fg1}a shows a two-color composite map produced using the {\it Spitzer} 4.5 $\mu$m (in red) and 3.6 $\mu$m (in turquoise) images, highlighting several areas dominated by extended 4.5 $\mu$m emission. Previously, Puga09 reported two blobs in S284-RE, which are labeled as bl-21 and bl-25 in the inset. 
The color-composite map is overlaid with the positions of Class~I YSOs (diamonds) and Class~II YSOs (squares) obtained from Puga09. 
Based on visual inspection, groups of YSOs can be seen, aligned in a trail, indicating the likely existence of a filamentary structure containing the blob bl-21 (see Section~\ref{sec:datanw} for more details). One can note that no YSOs are seen toward the blob bl-25. These two blobs are also evident in the {\it Herschel} image at 160 $\mu$m. Point-like sources detected in the {\it Herschel} image at 160 $\mu$m are found in areas, where YSOs are aligned in a trail.  

To further explore the extended 4.5 $\mu$m emission, Figure~\ref{fg1}b displays a ratio map of the 4.5 $\mu$m/3.6 $\mu$m emission, showing several bright areas. 
These bright areas are dominated by the 4.5 $\mu$m emission. In the literature, the use of this {\it Spitzer} ratio map, together with radio continuum emission, has provided valuable insights into molecular outflows and the impact of massive stars on their environments \citep[e.g.,][]{dewangan_2010, dewangan16,bhadari_2020}. Due to the similar point response function (PRF) of {\it Spitzer} 3.6 $\mu$m and 4.5 $\mu$m images, a ratio map can be produced by dividing the 4.5 $\mu$m image by the 3.6 $\mu$m image. The {\it Spitzer} 4.5 $\mu$m band includes a hydrogen recombination line (Br$\alpha$ at 4.05 $\mu$m) and strong molecular hydrogen (H$_{2}$) emission ($\nu$ = 0--0 $S$(9) at 4.693 $\mu$m), which are often excited by outflow shocks. The {\it Spitzer} 3.6 $\mu$m band, on the other hand, contains polycyclic aromatic hydrocarbon (PAH) emission at 3.3 $\mu$m. 

A 6-pixel median filter is first applied to the ratio map, followed by smoothing using a 3 $\times$ 3 pixel boxcar filter. The resulting ratio map is presented in Figure~\ref{fg1}b.
Given the absence of radio continuum emission, the bright regions likely correspond to H$_{2}$ features, while the dark areas trace PAH emission, suggesting the presence of warm dust \citep[see Figure 4 in][]{bhadari23}. In several areas, these bright regions (except bl-25) appear elongated, and are associated with YSOs, suggesting outflow activities. 
The dotted cyan box in Figure~\ref{fg1}b outlines the area covered by the {\it JWST} observations.  
The {\it Spitzer} ratio map is overlaid with the ALMA 1.38 mm continuum emission contours (in red, cyan, and magenta), highlighting the presence of embedded continuum sources hosting YSOs driving the observed outflows. Additionally, based on the coverage of the ALMA observations, three sub-regions---Rg1, Rg2, and Rg3---are marked in Figure~\ref{fg1}b. While ALMA data are available for Rg3, no {\it JWST} data are present for this sub-region. The blob bl-25 is not observed by neither the ALMA nor {\it JWST} facilities. Hence, our current work focuses on the Rg1, Rg2, and Rg3 subregions of S284-RE.

\subsection{Herschel Column density and dust temperature maps}
\label{sec:data2}
{\it Herschel} images at 160--500 $\mu$m allow us to generate column density ($N(\rm H_{2})$) map and dust temperature ($T_{\rm d}$) map. We used the {\tt hires} tool \citep{getsf_2022} to generate the $N(\mathrm{H}_2)$ and $T_{\rm d}$ maps with different resolutions ranging from 
$\sim$8$''$ to $\sim$36\rlap.{$''$}3 \citep[see more details in][]{Dewangan_2023,Dewangan_2024,Maity_23}. However, the 8$''$ image is not used in this work due to the presence of warm dust in the {\it Herschel} 70 $\mu$m image toward S284-RE.

\begin{figure*}
\center
\includegraphics[width=\textwidth]{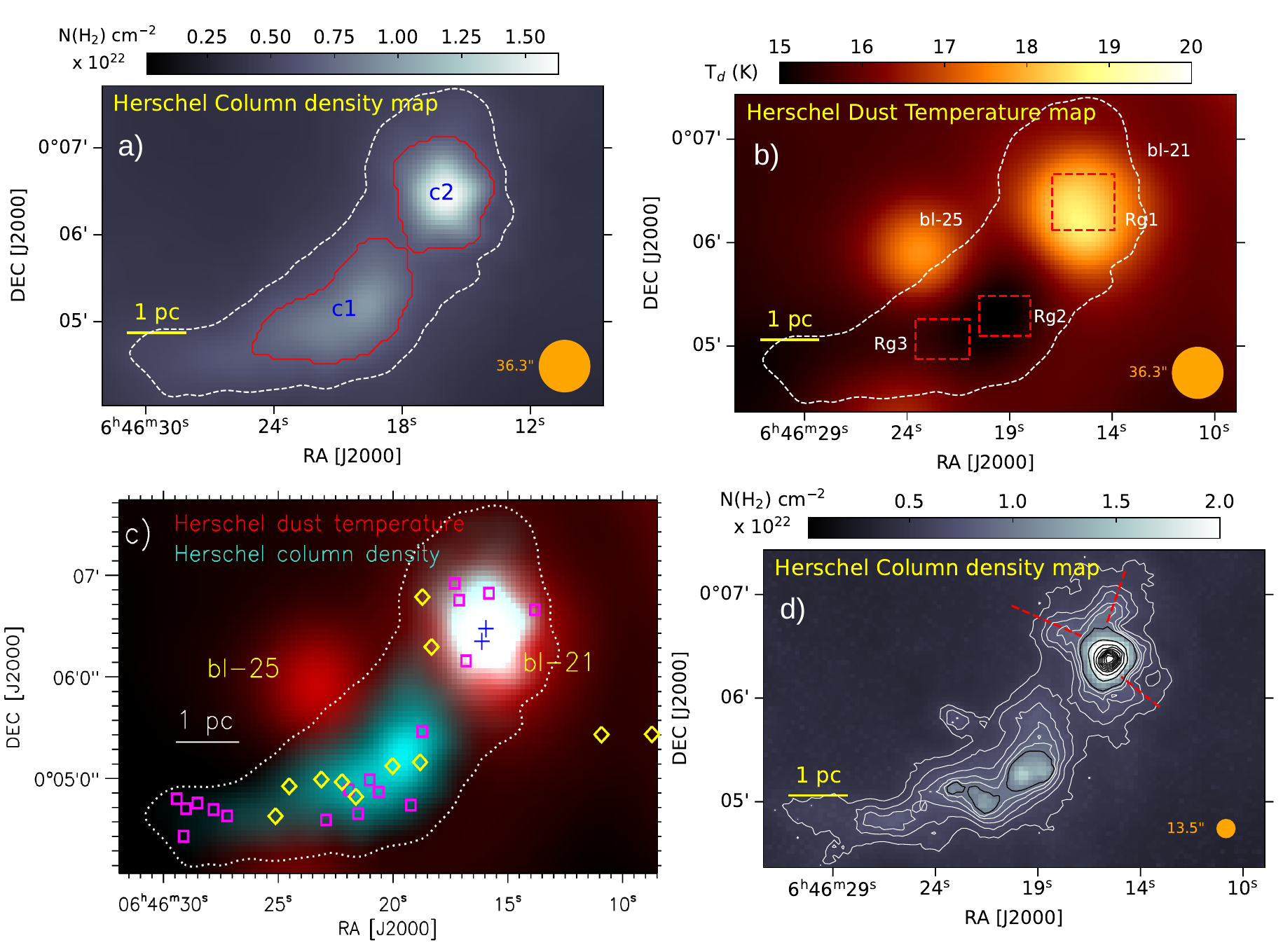}
\caption{a) {\it Herschel} column density ($N(\rm H_{2})$) map (resolution $\sim$36\rlap.{$''$}3) 
of an area highlighed by the solid box in Figure~\ref{fg1}a. 
The $N(\rm H_{2})$ contour (see dashed curve) is shown with a level of 5 $\times$ 10$^{21}$ cm$^{-2}$. 
The boundaries of two clumps (i.e., c1 and c2) traced in the $N(\rm H_{2})$ map 
are marked by red curves.
b) {\it Herschel} temperature ($T_{\rm d}$) map (resolution $\sim$36\rlap.{$''$}3). 
The areas presented in the ALMA continuum maps for Rg1, Rg2, and Rg3 are highlighted by dashed boxes 
(see Figures~\ref{fg7}a--\ref{fg7}c). The dashed contour is the same as in Figure~\ref{fg2}a. 
c) The panel shows a two-color composite map made using the $T_{\rm d}$ map (in red) and $N(\rm H_{2})$ map (in turquoise) overlaid with the 
positions of Class~I YSOs (diamonds) and Class~II YSOs (squares) obtained from Puga09. 
Two blobs (i.e., bl-21 and bl-25) studied in Puga09 are also labeled.   
d) {\it Herschel} column density map (resolution $\sim$13\rlap.{$''$}5) overlaid with $N(\rm H_{2})$ contours. The white contour levels are 5.5, 6.4, 7.3, 8.2, 9.1, and 10 $\times$ 10$^{21}$ cm$^{-2}$. The contours in black are at levels 1.0, 1.6, 2.3, 3.0, 3.6, 4.3, 5.0, 5.6, 6.3, and 7 $\times$ 10$^{22}$ cm$^{-2}$ highlighting the regions with higher $N(\rm H_{2})$. In each panel, the scale bar corresponding to 1 pc is shown at a distance of 5.0 kpc.}  
\label{fg2}
\end{figure*}

Figures~\ref{fg2}a and~\ref{fg2}b show the {\it Herschel} $N(\rm H_{2})$ and $T_{\rm d}$ maps (resolution$\sim$36\rlap.{$''$}3) of S284-RE, respectively. 
The range of the $N(\rm H_{2})$ is [0.32, 1.63] $\times$ 10$^{22}$ cm$^{-2}$, while $T_{\rm d}$ spans from 14.9 to 18.7~K.  
A filamentary structure is traced at the $N(\rm H_{2})$ value of $\sim$5 $\times$ 10$^{21}$ cm$^{-2}$ (see the dashed contour in Figures~\ref{fg2}a and~\ref{fg2}b). 
Rg1, Rg2, Rg3, bl-21, and bl-25 are also labeled in Figure~\ref{fg2}b.  
The higher value of $N(\rm H_{2})$ is found toward Rg1 (or bl-21) compared to Rg2 and Rg3.  
We used the {\tt RadFil}\footnote[3]{https://github.com/catherinezucker/radfil?tab=readme-ov-file} algorithm \citep{Zucker_2018}, which uses medial axis skeletonization on the filament mask (which is provided by the user) to calculate the spine and the length of the filament. The length of the S284-RE filament is determined to be $\sim$8.5 pc. 
A variation in $T_{\rm d}$ is also evident along the length of the filamentary structure (see the dashed contour in Figure~\ref{fg2}b). 
We find the presence of warm dust emission with $T_{\rm d}$ $\sim$17--18~K toward Rg1 (or bl-21) and bl-25, while Rg2 and Rg3 are seen with cold dust emission (i.e., $T_{\rm d}$ $\sim$15.0--15.7~K). Figure~\ref{fg2}c presents a two-color composite map made using the $T_{\rm d}$ (in red) and $N(\rm H_{2})$ maps (in turquoise), displaying the embedded structure in S284-RE. The map shows a higher concentration of YSOs in the regions with cold dust emission (i.e., Rg2 and Rg3).

We calculated the mass of the filament using the $N(\rm H_{2})$ map (resolution $\sim$ 36\rlap.{$''$}3) and the following equation \citep[e.g.,][]{Dewangan_2017}:
\begin{equation}
	M_{\rm clump}= \mu_{\rm H_{2}}~m_{\rm H}~ a_{\rm pixel}~\Sigma N(\mathrm H_2),
\label{eq1}
\end{equation}
 where $\mu_{\rm H_{2}}$ is the  ${\rm H_2}$ mean molecular weight (assumed to be 2.8), $m_{\rm H}$ is the mass of the hydrogen atom, $a_{\rm pixel}$ is the physical area subtended by 1 pixel in cm$^2$, and $\Sigma N(\mathrm H_2)$ is the integrated column density. 
For obtaining $\Sigma N(\mathrm H_2)$, we integrated the $N(\rm H_{2})$ above the contour level of 5 $\times$ 10$^{21}$ cm$^{-2}$.  
The mass of the filament is computed to be $\sim$2402 $M_{\odot}$. The typical error in this mass estimate is $\sim$50$\%$, taking into account the uncertainities in the estimation of $N(\rm H_{2})$ and $a_{\rm pixel}$, which are $\sim$25$\%$ and $\sim$18$\%$, respectively taking into account the uncertainity in the estimation of distance as $\sim$9$\%$.

To assess the stability of a filament, one needs to calculate the observed line-mass $M_\mathrm{l,obs}$ and the critical line mass $M_\mathrm{l,crit}$ of the filament \citep{2010A&A...518L.102A}. The $M_\mathrm{l,crit}$ can be estimated using the equation
16$\times (\frac{T}{10~{\rm K}})$  $M_{\odot}$ pc$^{-1}$ \citep{2014prpl.conf...27A}. The filament can be identified as thermally supercritical or subcritical based on its line-mass. A thermally supercritical filament has a line-mass exceeding the critical value ($M_\mathrm{l,obs} > M_\mathrm{l,crit}$), making it susceptible to gravitational collapse \citep[e.g.,][]{bhadari23}. In contrast, a subcritical filament has a line-mass below the critical threshold ($M_\mathrm{l,obs} < M_\mathrm{l,crit}$), suggesting it remains stable against collapse.
Considering the observed values of $M$ ($\sim$2402 $M_{\odot}$) and length ($\sim$8.5 pc) of the filament, 
the value of $M_\mathrm{l,obs}$ is determined to be $\sim$282 M$_{\odot}$ pc$^{-1}$. As inferred from the {\it Herschel} dust temperature map, we have used an average temperature of $\sim$18 K. The $M_\mathrm{l,crit}$ of the filament is estimated to be $\sim$29 $M_{\odot}$ pc$^{-1}$. 
It implies that $M_\mathrm{l,crit}$ is less than $M_\mathrm{l,obs}$, indicating that the filament is thermally supercritical.

We also employed the {\tt astrodendro}\footnote[4]{https://dendrograms.readthedocs.io/en/stable/} package in Python to identify clumps in the filament using the $N(\rm H_2)$ map (resolution $\sim$ 36\rlap.{$''$}3). The input parameters for source detection were: 1) {\it min\textunderscore value}, the minimum flux value for detection, 2) {\it min\textunderscore delta}, the minimum peak flux difference to define an independent source, and 3) {\it min\textunderscore pix}, the minimum number of pixels required to define a source. We set the input parameters as, {\it min\textunderscore value} = 3$\sigma$, {\it min\textunderscore delta} =  3$\sigma$, and {\it min\textunderscore pix} = 2 $\times$ the beam area \citep[see more details in][]{Rosolowsky08,bhadari23}. The boundaries of the two identified clumps, c1 and c2, are shown in Figure~\ref{fg2}a. The masses of clumps c1 and c2 are determined to be $\sim$560 and $\sim$580 $M_{\odot}$, respectively.

In Figure~\ref{fg2}d, we have presented the {\it Herschel} $N(\rm H_{2})$ map at a resolution of $\sim$13\rlap.{$''$}5, allowing to infer sub-structures toward the filament. High resolution $N(\rm H_{2})$ map shows more fragments in the filament, which are not seen in Figure~\ref{fg2}a. Additionally, visual inspection reveals at least three elongated emission features (marked by red lines) surrounding clump c2, hinting at a possible hub-filament system around the clump. 
\subsection{Clustering analysis of YSOs in S284-RE}
\label{sec:datanw}
\begin{figure*}
    \centering
    \includegraphics[width=\textwidth]{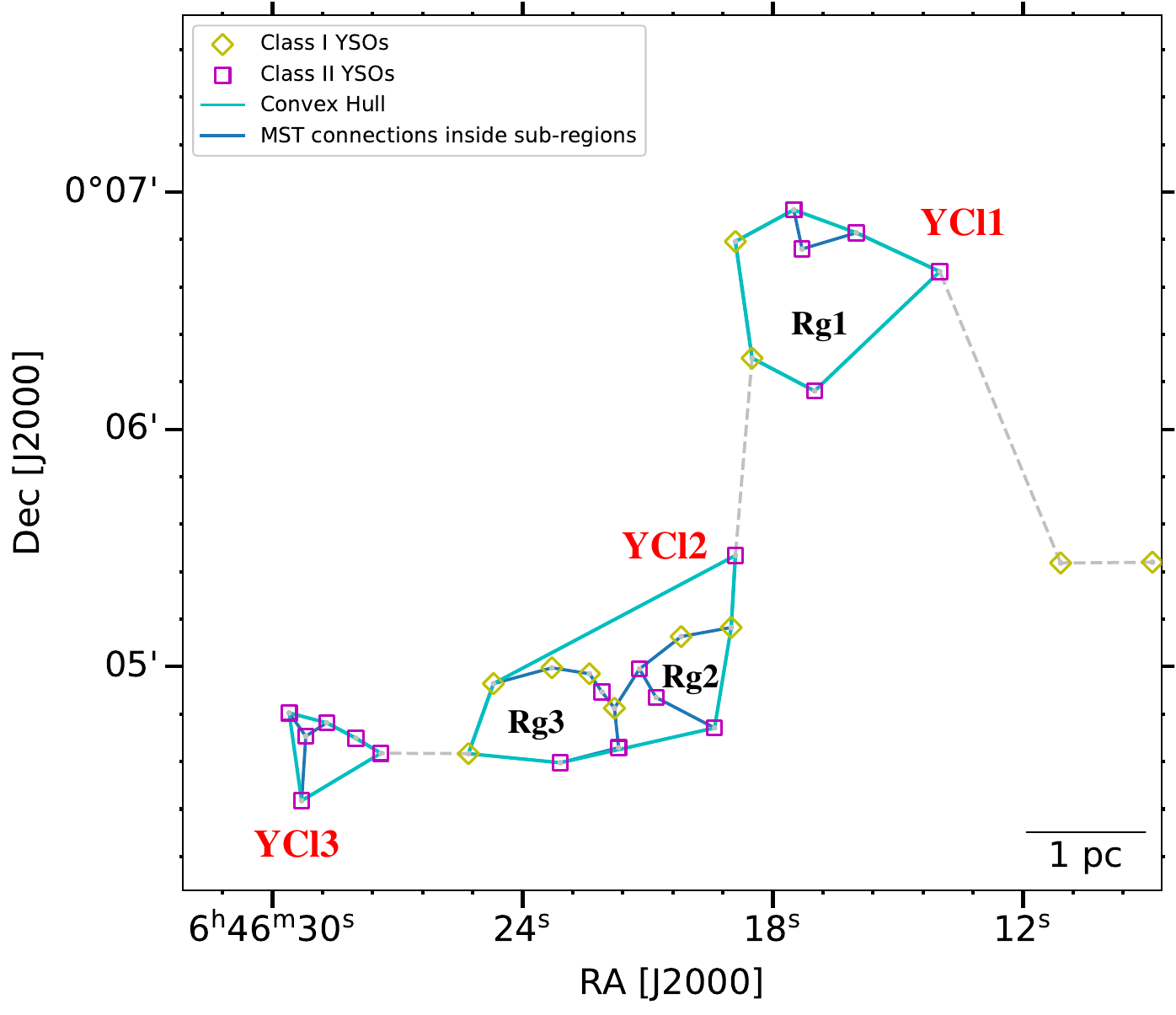}
     \caption{The panel shows the MST connections in YCl1, YCl2, and YCl3. 
     The locations of Class~I YSOs (yellow diamonds) and Class~II YSOs (magenta squares) are marked. 
     The extracted clusters YCl1, YCl2, and YCl3 are also enclosed by the \emph{Convex hulls} using 
     cyan line segments. The MST connections inside each cluster are shown with blue line segments. Sub-regions (i.e.,Rg1, Rg2, and Rg3) are also highlighted in the figure.}   
    \label{fig:mst}
\end{figure*}

In this section, we applied an empirical technique, \emph{`Minimal Spanning Tree'} (MST; \citealt{2009ApJS..184...18G,2014MNRAS.439.1765K}), which is used to isolate the clustering/sub-clustering of YSOs with neutrality \citep{2004MNRAS.348..589C, 2016AJ....151..126S, 2020ApJ...891...81P,2024AJ....168...98V}. Figure~\ref{fig:mst} presents the extracted MST, highlighting sub-regions Rg1, Rg2, and Rg3, as well as the detected clusters YCl1, YCl2, and YCl3. The cluster YCl1 is linked with Rg1, while the cluster YCl2 is associated with Rg2 and Rg3. Note that ALMA and {\it JWST} maps are not available toward YCl3.  
In this analysis, we determined the critical branch length using the histogram of the MST branch lengths \citep[see][for more details]{2009ApJS..184...18G,2016AJ....151..126S,2020ApJ...891...81P}. The distribution of YSOs in the regions/sub-regions is often asymmetric. Therefore, we used the \emph{convex hull/Qhull}\footnote[5]{\emph{Convex hull} is a polygon that encloses all the points or objects within a given sub-region, where the internal angles between any two consecutive sides are less than 180$^{\circ}$.} method to estimate the cluster area $A_{\rm cluster}$, which is normalized by a geometric factor \citep{2006A&A...449..151S,2016AJ....151..126S,2020MNRAS.498.2309S,2023ApJ...953..145V}, as defined by the following formula:
\begin{equation}
 A_{\rm cluster}=\frac{A_{\rm hull}}{1-\frac{n_{\rm hull}}{n_{\rm total}}},
\end{equation}
where, $A_{\rm hull}$ is the area of the $Q_{hull}$, $n_{\rm hull}$ is the total number of vertices on the hull, and $n_{\rm total}$ is the total number of points/objects inside the $Q_{hull}$. 
The radius of a circle with an area equal to $A_{\rm cluster}$ is defined as the radius of the sub-region $R_{\rm cluster}$. The extracted sub-regions are enclosed by \emph{Convex hulls}, shown by the cyan line segments 
in Figure~\ref{fig:mst}. We also calculated $R_{\rm circ}$, which is defined as half of the maximum distance between two hull objects, and the aspect ratio ($\frac{R^2_{\rm circ}}{R^2_{\rm cluster}}$) for the cluster  YCl1, YCl2, and YCl3. These parameters are listed in Table \ref{tab:mst_parameter}.

\subsection{{\it JWST} images}
\label{sec:data4}
\begin{figure*}
\center
\includegraphics[width=12.5cm]{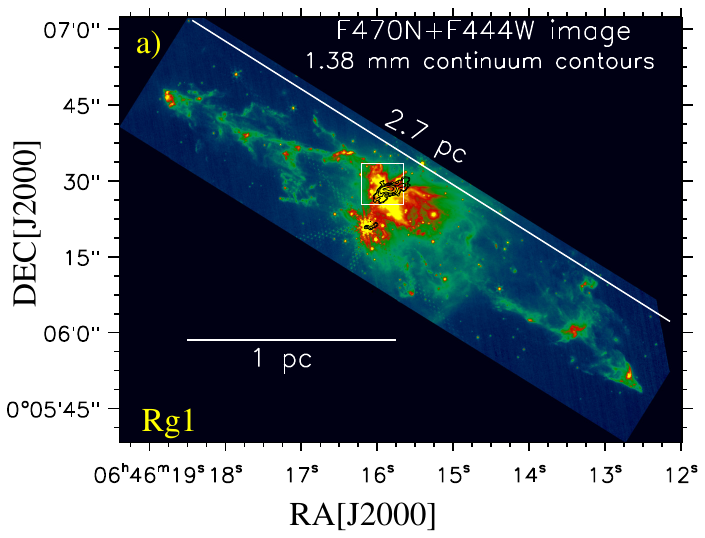}
\includegraphics[width=10.5cm]{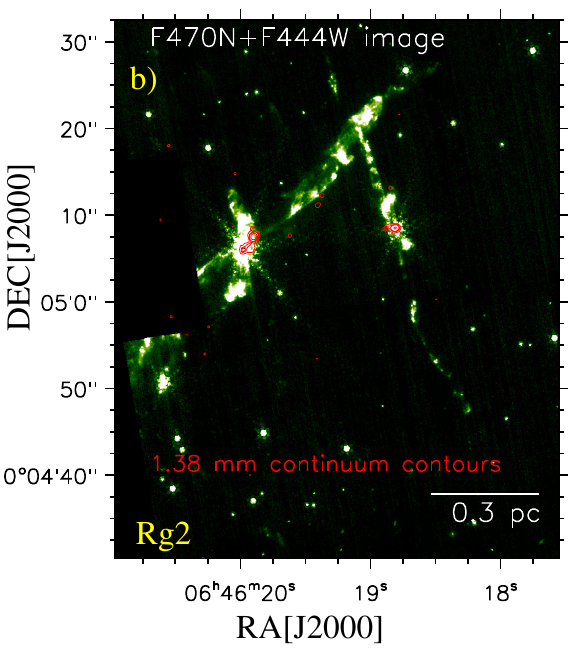}
\caption{a) Overlay of the ALMA 1.38 mm continuum emission contours on the {\it JWST} F470N+F444W image 
toward Rg1 (see yellow dot-dashed box in Figure~\ref{fg1}b). 
The black contours are shown with the levels of 0.9, 1.2, 2.0, 3.0 and 7.0 mJy beam$^{-1}$. 
b) Overlay of the ALMA 1.38 mm continuum emission contours on 
the {\it JWST} F470N+F444W image toward Rg2 (see magenta dot-dashed box in  Figure~\ref{fg1}b). 
The red contours are shown with the levels of 0.43, 0.8, 1.0, and 2.0 mJy beam$^{-1}$. Each scale bar represents the physical scale at the distance of 5.0 kpc.}
\label{fg4}
\end{figure*}

In the direction of S284-RE, the {\it Spitzer} images and their ratio map show the presence of molecular outflows driven by YSOs. 
Figures~\ref{fg4}a and~\ref{fg4}b display the {\it JWST} F470N+F444W images of sub-regions Rg1 and Rg2, highlighting the emission associated with outflows traced by H$_2$ emission (see dot-dashed boxes in Figure~\ref{fg1}b). These images are overlaid with contours of ALMA 1.38 mm continuum emission. % to highlight the possible driving sources.}
In sub-region Rg1, the {\it JWST} F470N+F444W image clearly shows multiple outflows, with one particulary prominent bipolar outflow extending $\sim$2.7 pc (see Figure~\ref{fg4}a). 
The ALMA 1.38 mm continuum emission contours are detected in the central part of this outflow, 
where several embedded YSOs seem to be located. 
In Figure~\ref{fg4}b, we find at least three outflows in the {\it JWST} F470N+F444W image of sub-region Rg2. 
The peaks of the ALMA 1.38 mm continuum emission can be utilized to infer the positions of the driving sources for these outflows.

\begin{figure*}
\center
\includegraphics[width=\textwidth]{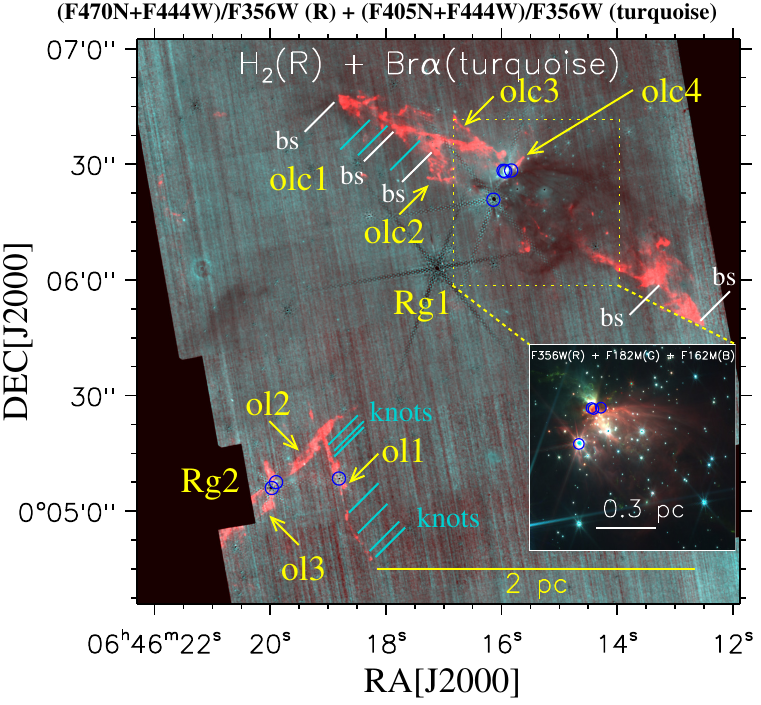}
\caption{The panel shows a two-color composite map made using the {\it JWST} 
ratio (F470N+F444W)/F356W map or H$_{2}$ emission (in red) and {\it JWST} ratio (F405N+F444W)/F356W map or Br-$\alpha$ emission (in turquoise). An inset on the bottom right displays a zoomed-in view of the 
selected area indicated by the dotted box in Figure~\ref{fg5}. The inset is a three-color composite map made using the F356W (in red), F182M (in green), and F162M (in blue) images. At least four H$_{2}$ outflows (olc1--olc4) are identified in Rg1, while three H$_{2}$ outflows (ol1--ol3) are observed in Rg2. The driving sources of the outflows which are identified using the JWST NIRCAM images and ALMA continuum sources are 
marked by blue circles (see also the inset). H$_{2}$ knots and bow-shock (bs) are indicated.  
Each scale bar is derived at a distance of 5.0 kpc.}  
\label{fg5}
\end{figure*}

To further explore these sub-regions, we have produced the ratio maps using the {\it JWST} images, which are (F470N+F444W)/F356W and (F405N+F444W)/F356W maps. Both these ratio maps are processed using median filtering with a 5-pixel width, followed by smoothing with a 3 $\times$ 3 
pixel boxcar algorithm. In Figure~\ref{fg5}, we present a two-color composite map made using (F470N+F444W)/F356W map (in red) and (F405N+F444W)/F356W map (in turquoise). 
The bright features in (F470N+F444W)/F356W and (F405N+F444W)/F356W maps trace the H$_{2}$ emission at 4.693 $\mu$m (in red) and Br-$\alpha$ emission at 4.05 $\mu$m (in blue), respectively. 
On the other hand, the dark features in both these {\it JWST} ratio maps appear to depict the PAH emission at 3.3 $\mu$m \citep[see][]{dewangan24b}. 
In Rg1, we highlight at least four H$_{2}$ outflows (olc1--olc4) and driving sources (see blue circles; see Figure~\ref{fg5}). Similarly, in Rg2, we mark three H$_{2}$ outflows (ol1--ol3) and driving sources (see blue circles; see Figure~\ref{fg5}). We also identify distinct bow-shocks and knots in H$_{2}$ emission associated with the outflow olc1 (see Figure~\ref{fg5}). In the direction of ol1, several H$_{2}$ knots are clearly visible. Considering the spatial geometry of H$_{2}$ emission, 1.38 mm continuum peak emission, and previously known locations of YSOs, these driving sources are identified. Note that the central part of the outflow olc1 in Rg1 seems to host several embedded protostars and is dominated by the 3.3 $\mu$m emission, particularly on the southern side of the outflow olc4. The inset in Figure~\ref{fg5} displays a three-color composite map made using the F356W (in red), F182M (in green), and F162M (in blue) images overlaid with the positions of the outflow driving sources. In addition to the extended emission observed in the F356W image, a small-scale nebulous feature (extent $\sim$14000 AU $\times$ 9500 AU), having an almost elliptical morphology, is prominently evident on the northern side of the outflow olc4 in the F162M and F182M images. 

\begin{figure*}
\center
\includegraphics[width=\textwidth]{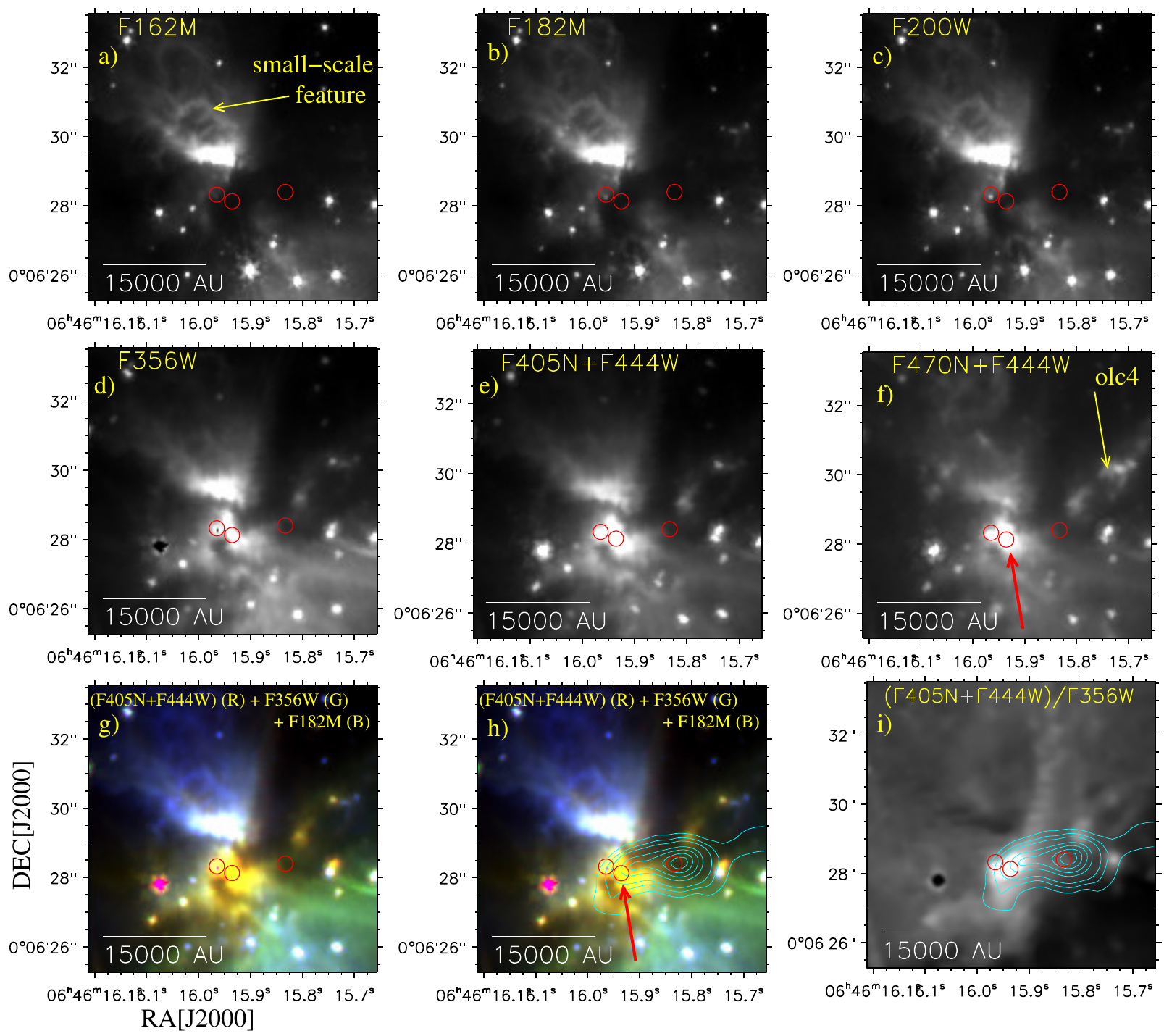}
\caption{a--f) Multi-wavelength view of an area highlighted by the solid box in 
Figure~\ref{fg4}a. g) The panel shows a three-color composite map made 
using the F405N+F444W (in red), F356W (in green), and F182M (in blue) images. 
h) Overlay of the ALMA 1.38 mm continuum contours on the {\it JWST} three-color composite map. 
The composite map is the same as Figure~\ref{fg6}g. The continuum contour levels are 2, 3, 4, 6, 7.2, 8.5, and 9.5  mJy beam $^{-1}$. 
The red arrow highlights the location of the massive protostar candidate. 
i) The {\it JWST} ratio (F405N+F444W)/F356W map overlaid with the ALMA 1.38 mm continuum emisson contours (see Figure~\ref{fg6}h). 
In each panel, red circles highlight embedded point-like sources detected in the {\it JWST} images and/or the ALMA continuum maps. In each panel, the scale bar corresponding to 15000 AU is derived at a distance of 5.0 kpc.} 
\label{fg6}
\end{figure*}

In order to further probe the small-scale feature and the driving source of the bipolar outflow olc1, a zoomed-in view of an area indicated by the solid box in Figure~\ref{fg4}a is presented using the {\it JWST} F162M, F182M, F200W, F356W, (F405N+F444W), (F470N+F444W) images in Figures~\ref{fg6}a--\ref{fg6}f, respectively. 
Figure~\ref{fg6}g displays a three-color composite map, with the F405N+F444W (in red), F356W (in green), and F182M (in blue) images. Figure~\ref{fg6}h shows the same composite as Figure~\ref{fg6}g, but with the ALMA 1.38 mm continuum contours overlaid. 
Figure~\ref{fg6}i presents the overlay of the ALMA 1.38 mm continuum contours on the {\it JWST} ratio (F405N+F444W)/F356W map. 
In Figures~\ref{fg6}a--\ref{fg6}i, red cirlces show the location of point sources identified using {\it JWST} and ALMA continuum images.
The powering source ($\alpha_{2000}$ = 06$^{h}$46$^{m}$15\rlap.$^{s}$94; $\delta_{2000}$ = 00$\degr$06$'$28\rlap.{$''$}13; see the red arrow in Figure~\ref{fg6}h) of the outflow olc1, seems to be associated with the ionized emission or Br-$\alpha$ emission traced in the {\it JWST} ratio (F405N+F444W)/F356W map and the 1.38 mm continuum emission. 
Another ALMA continuum peak is detected, which seems to host the driving source of the outflow olc4. 
The small-scale feature is more pronounced  in the {\it JWST} SW F162M, F182M, and F200W images, while the southern side of this feature is more prominent 
in the {\it JWST} LW F356W, F405N, and F470N images, revealing deeply embedded region associated with the 1.38 mm continuum emission. 
\subsection{ALMA continuum sources}
\label{sec:data5}
\begin{figure*}
\center
\includegraphics[width=18cm]{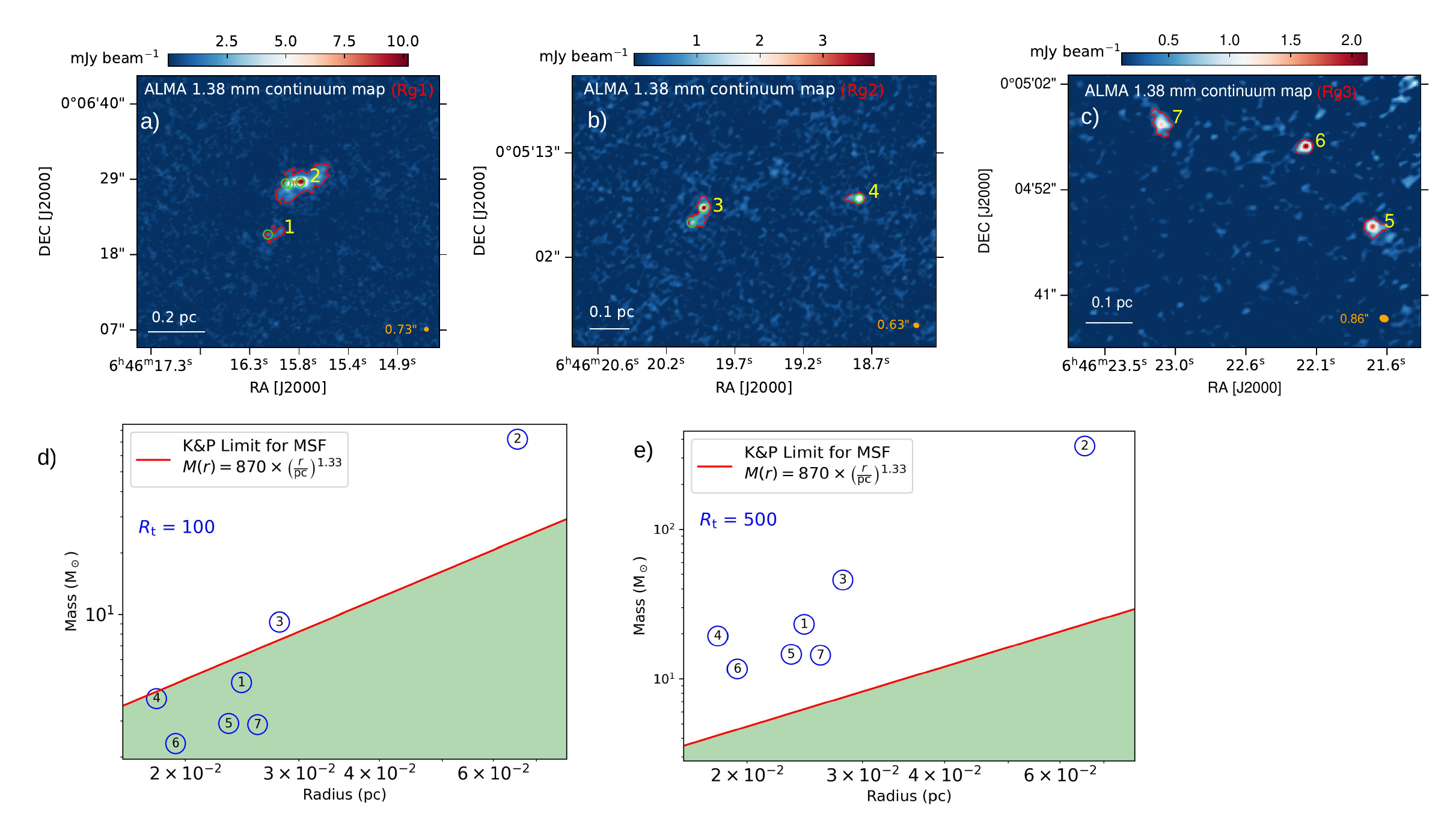}
\caption{ALMA 1.38 mm continuum emission map of a) Rg1, b) Rg2, and c) Rg3 overlaid with red contours highlighting the detected continuum sources (see the dotted box in 
Figure~\ref{fg2}a). In panels ``a--c'', the continuum contour is plotted at 3$\sigma$, the driving sources are highlighted using open circles (in green), and the scale bar is derived at a distance of 5.0 kpc. d) The panel shows the $M$--$R_{\rm eff}$ plot of the continuum sources detected in the ALMA 1.38 mm continuum emission map toward Rg1, Rg2, and Rg3 (see dotted boxes in Figure~\ref{fg2}a). 
The region (in white) above the red line corresponds to the condition for massive star formation \citep[from][]{kauffmann10}. 
Masses are computed for R$_{t}$ = 100. e) Same as Figure~\ref{fg7}d, but masses are determined for R$_{t}$ = 500.} 
\label{fg7}
\end{figure*}

The {\tt astrodendro} python package is employed to identify continuum sources in the ALMA 1.38 mm continuum map toward Rg1, Rg2, and Rg3 regions (see red, magenta, and cyan contours in Figure~\ref{fg1}b). We set the input parameters as, {\it min\textunderscore value} = 3$\sigma$, {\it min\textunderscore delta} =  3$\sigma$, and {\it min\textunderscore pix} = 2 $\times$ the beam area. The detected sources toward Rg1, Rg2, and Rg3 are highlighted using red contours in Figure~\ref{fg7}a,~\ref{fg7}b, and~\ref{fg7}c, respectively. A total of 7 continuum sources are detected in the ALMA continuum maps. We calculated the mass of each continuum source using the following equation \citep{hildebrand83}:
\begin{equation}
M=\frac{F_\nu~d^2~R_{\rm t}}{B_{(\nu,T_{\rm d})}~\kappa_{\nu}},
\end{equation}
where $F_\nu$ is the integrated flux, $R_{\rm t}$ is the gas to dust ratio, $B_{(\nu,T_{\rm d})}$ is the Planck function at temperature $T_{\rm d}$, and $\kappa_\nu$ is the dust-absorption coefficient at frequency $\nu$. The $\kappa_\nu$ is estimated using, $\kappa_{\nu}=\left(\frac{\nu}{\rm 1.2 THz}\right)^{1.5}$ \citep{hildebrand83}. For our calculations, we adopted $d$ = 5.0 kpc and $R_{\rm t}$ = 100. 
However, since S284-RE is located in a low-metallicity region \citep[i.e., Z $\sim$0.2 Z$_{\odot}$; see][and references therein]{kalari15}, we have also used a value of \(R_{\rm t}\) = 500 (considering the relation for between Z and R$_{\rm t}$, as R$_{\rm t}$ $\sim$1/Z \citep{Remy_2014})
Here, we assume that the S284-RE region has the same metallicity as the pre-main sequence (PMS) stars in the S284 H\,{\sc ii} region. However, this assumption could be incorrect if some physical processes have led to lower metallicity in the PMS stars of the S284 H\,{\sc ii} region. To account for this possibility, we calculated the masses of the detected continuum sources using both \(R_{\rm t} = 100\) and \(R_{\rm t} = 500\). 
The effective radius, $R_{\rm eff}$, of the continuum sources is determined using the exact area (A) covered by the source, where $R_{\rm eff}=\sqrt{\frac{A}{\pi}}$ (assuming circular geometry for the source). 
The surface density of the sources is then calculated using the relation $\Sigma=\frac{M}{\pi R_{\rm eff}^{2}}$ \citep{Colombo_19}. Subsequently, the volume density is determined using the equation $n({\rm H_2})=\frac{3M}{4\pi\mu m_p R_{\rm eff}^3}$ \citep{Rigby_19}. The values of $M$, $R_{\rm eff}$, $\Sigma$, and $n({\rm H_2})$ for all the sources are listed in Table~\ref{tab:alma_params}. The typical error in the estimated mass values is $\sim$30 $\%$, considering the uncertainities in $F_{\nu}$, $d$, $R_{\rm t}$ (i.e., 100) and $T_{\rm d}$ as $\sim$15$\%$, $\sim$9$\%$ (Puga09), $\sim$23$\%$ \citep{Sanhueza_2017}, and $\sim$5$\%$ respectively. However, the uncertainty in $R_{\rm t}$ = 500 can be significant, leading to a mass estimate error of at least $\sim$50$\%$.
 The error in $R_{\rm eff}$ primarily arises from the uncertainty in the distance estimation, which is $\sim$9$\%$. This uncertainty propagates to an error of approximately 18$\%$ in  $R_{\rm eff}$. However, this represents a lower limit. The actual error in $R_{\rm eff}$, could be significantly larger, primarily due to our assumption of the source having a circular shape, which may not accurately reflect its true geometry.
%due to our assumption of a circular shape for the source, }

%However, the uncertainity in $R_{\rm t}$ = 500 can have large uncertainities, hence the error calculated using $R_{\rm t}$ = 500 can be at least $\sim$50 $\%$.

%=====================================================

\begin{table*}[ht]
    \centering
    \caption{Physical parameters of the continuum sources identified in the ALMA 1.38 mm continuum map using {\it astrodendro}.}
    \label{tab:alma_params}
    \begin{tabular}{cccccccccc}
        \hline
        Source ID & RA (hh:mm:ss) & DEC (dd:mm:ss)  & $R_{\rm eff}$ (pc)  &  \multicolumn{2}{c}{$M$ ($M_\odot$)} &    \multicolumn{2}{c}{ $\Sigma$ (g cm$^{-2}$)} &    \multicolumn{2}{c}{$n({\rm H_{2}})$ ($\times$ 10$^{6}$ g cm$^{-3}$)} \\
        \vspace{0.1 cm}
        ~&~~&~&~&\(R_{\rm t}\) = 500&\(R_{\rm t}\) = 100&\(R_{\rm t}\) = 500&\(R_{\rm t}\) = 100&\(R_{\rm t}\) = 500&\(R_{\rm t}\) = 100\\
        \hline
        1 &    06:46:16.08 & 00:06:21.01 &   0.024 &  	23.12 &	  4.62 &	     2.56 &    0.51&	     5.69&   1.12\\
        2 &    06:46:15.83 & 00:06:28.32 &   0.065 &    360.36 &	  72.07&	     5.56 &    1.11&	     4.54&   0.90\\
        3 &    06:46:19.91 & 00:05:07.10 &   0.027 &  	45.73 &	  9.15&	     3.86 &    0.77&	     7.36&   1.47\\
        4 &    06:46:18.82 & 00:05:08.52 &   0.018 &  	19.29 &	  3.86&	     3.92 &    0.78&	     11.60&  2.31\\
        5 &    06:46:21.69 & 00:04:47.75 &   0.023 &  	14.56 &	  2.91&	     1.76 &    0.35&	     4.03&   0.80\\
        6 &    06:46:22.16 & 00:04:55.96 &   0.019 &  	11.61 &	  2.32&	     2.06 &    0.41&	     5.68&   1.13\\
        7 &    06:46:23.12 & 00:04:58.38 &   0.025 &  	14.37 &	  2.87&	     1.42 &    0.28&	     2.92&   0.58\\
        \hline
    \end{tabular}
\end{table*}

%===================================================

To identify potential continuum sources hosting massive stars, we have plotted $M$ versus $R_{\rm eff}$ for all the detected continuum sources 
(see Figures~\ref{fg7}a--\ref{fg7}c). Figures~\ref{fg7}d and~\ref{fg7}e display the $M$ versus $R_{\rm eff}$ plot for the masses calculated considering \(R_{\rm t} = 100\) and \(R_{\rm t} = 500\), respectively.
 The red line in the plot represents criterion for massive star formation, defined as 870 $M_{\odot}\times(\frac{R}{\rm pc})^{1.33}$ \citep[from][]{kauffmann10}. Continnum sources lying above this threshold are considered favorable for the formation of massive stars. For \(R_{\rm t} = 100\), only continuum sources \#2 and \#3 satisfy this criterion for massive star formation. However, when \(R_{\rm t} = 500\), all seven continuum sources become potential candidates for massive star formation.

\begin{table*}
 \tiny
%\scriptsize
\small
\setlength{\tabcolsep}{0.15in}
%\small
%\scriptsize
%\setlength{\tabcolsep}{0.18in}
\centering
\caption{Properties of cluster YCl1, YCl2, and YCl3 based on the MST Analysis. The number of enclosed Class~I YSOs (N$^{a}$) is listed in column 2; fraction of Class~I YSOs (fraction$^b$) in column 3; $R_{\rm cluster}$ is mentioned in column 4, $R_{\rm circ}$ in column 5; aspect ratio in column 6; the mean MST branch length in column 7; Q parameter in column 8; total molecular mass in column 9; $\lambda_J$ in column 10; and the SFE is mentioned in column 11. }
%Here, $a$ represents the number of enclosed YSOs and $b$ is ratio of the enclosed Class~I YSOs to the total (= Class~I + Class~II) YSOs.}
\label{tab:mst_parameter}
\begin{tabular}{lccccccccccccccccccccccr}
\hline 
 Cluster & N$^a$ & fraction$^b$ &  $R_{\rm cluster}$ & $R_{\rm circ}$ & Aspect & MST & $Q$ & Molecular & $\lambda_{\rm J}$ & SFE\\
  & & & (pc) & (pc) & Ratio & (pc) &  & mass ($M_\odot$) & (pc) & (\%)\\
    \hline
    YCl1         & 7   & 0.28 & 1.61 & 0.90 & 0.31 & 1.79 &  0.27 &  24.84 & 4.48 & 17.45\\
    YCl2         & 14  & 0.5  & 0.95 & 1.31 & 1.89 & 1.17 &  0.25 & 136.35 & 0.92 &  6.67\\
    YCl3         & 6   & 0.0  & 0.49 & 0.71 & 0.64 & 0.84 & 0.28 &  25.53 & 0.76 & 14.99\\
\hline          
\end{tabular}
\end{table*} 

\section{Discussion}
\label{sec:disc}
The S284-RE sub-region is located away from the edges of the extended S284 complex but remains associated with it (see Figure~\ref{fg1}a). As pointed out by Puga09, S284-RE appears to be influenced by the feedback from massive stars. Based on new observational data sets and previously reported results, new outcomes are derived and their implications are discussed in this section.
%
%\subsection{Investigating the thermally supercritical filament and associated sub-regions}
%\label{sec:discs1}
%
%In this section, we discuss the implications of the results derived using {\it Herschel}, {\it Spitzer}, ALMA and {\it Planck} facilities toward S284-RE. 
%
\subsection{Detection of a Thermally Supercritical Filament in S284-RE}
\label{sec:discs1a}
Puga09 observed that the distribution of YSOs in the S284-RE sub-region aligns along a trail (see Figure~\ref{fg1}a), which could be indicative of a underlying embedded structure. Generally, the spatial distribution of YSOs in a given star-forming region reflects the morphology of the dense areas of the parental molecular cloud \citep[e.g.,][]{chavarria08,guieu09,gutermuth08,gutermuth09,billot10,bhadari_2020,bhadari22}. In the case of S284-RE, the alignment of YSOs also seems to hint at the presence of a filamentary structure (see Figure~\ref{fg1}a). 
Using the {\it Herschel} column density map, we have confirmed the presence of a filament (length $\sim$8.5 pc), classified as thermally 
supercritical (see Section~\ref{sec:data2}). In the case of thermally supercritical filament, gravitational pressure dominates over thermal pressure within the filament \citep{2010A&A...518L.102A}, leading to radial collapse perpendicular to the filament’s major axis. This radial collapse of the filament is accompanied by the fragmentation of the filamentary cloud, ultimately resulting in the formation of YSOs. These fragments are clearly observed in the high-resolution N(H$_2$) map (indicated by the black contours in Figure~\ref{fg2}d), and are consistent with the spatial distribution of YSOs in the filament. 
Studies also suggest that thermally supercritical filaments undergo gravitational contraction, with their $M_\mathrm{l,obs}$ growing over time as they gather material from their surroundings \citep[e.g.,][]{Arzoumanian_2013}. In S284-RE, the filament also exhibits a temperature gradient and contains {\it Herschel} clumps, as shown in Figures~\ref{fg2}a--\ref{fg2}d. The presence of clusters of YSOs, signs of outflow activity, and ALMA continuum sources along the filament suggests active star formation (see Section~\ref{sec:datanw} and also Figure~\ref{fg1}b), supporting the radial collapse and fragmentation scenario in the cloud \citep[e.g.,][]{Arzoumanian_2013,Palmeirim_2013,Ncox_2016}. 
The filamentary structure in S284-RE might also have been influenced by the expansion of the S284 H\,{\sc ii} region and the associated stellar winds.

\begin{figure*}
\center
\includegraphics[width=16.5cm]{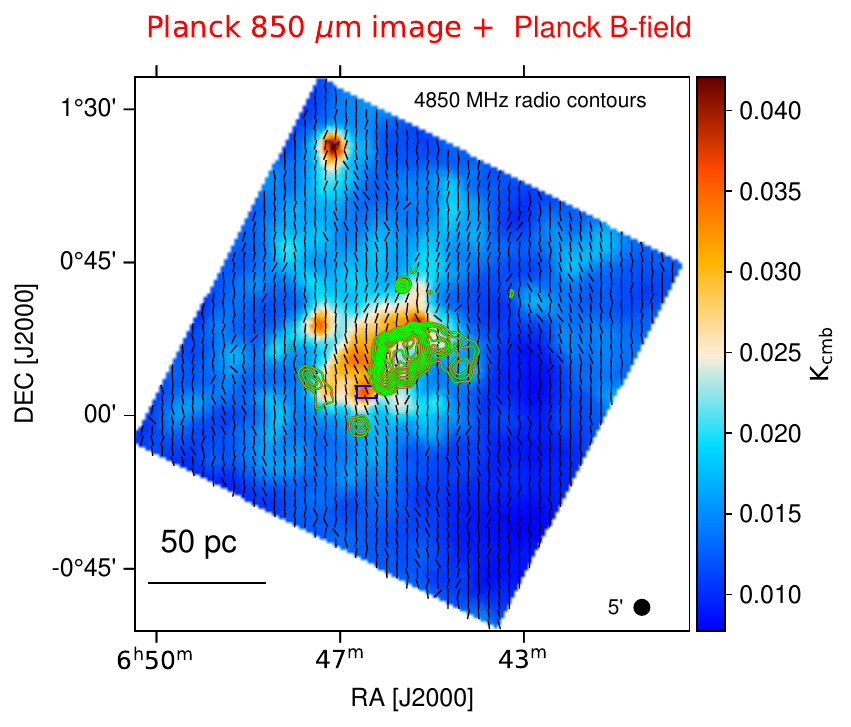}
\caption{Planck B-field segments (in black) overlaid on the Planck 850 $\mu$m (beamsize $\sim$5$'$) image in the units of thermodynamic temperature (K$_{\rm cmb}$). Each segment represents the average B-field orientation for an area of 9 arcmin$^2$. The 4850 MHz radio continuum contour levels (in green) are at levels 0.23 Jy beam$^{-1}$ $\times$ (0.15, 0.2, 0.3, 0.35, 0.4, 0.45, 0.5, 0.6, 0.7, 0.8, 0.9, 0.98). The blue box indicates the area highlighted in Figure~\ref{fg1}a. The scale bar is derived at a distance of 5 kpc. The Planck beam size is shown in the bottom right corner of the image.} 
\label{fg3}
\end{figure*}

Figure~\ref{fg3} displays the segments showing a large-scale B-field morphology inferred from the {\it Planck} 850~$\mu$m polarization observations toward the S284 H\,{\sc ii} region (see the 4850 MHz continuum emission contours in Figure~\ref{fg3}). The B-field lines appear to bend around the edge of the H\,{\sc ii} region. While, the B-field lines seem to be perpendicular to the filament located in S284-RE (see the blue box in Figure~\ref{fg3}). This is a commonly observed signature in the {\it Planck} observations toward the filaments with high column density \citep[log($N(\rm H_2)$) $\sim$ 21.5 cm$^{-2}$;][]{2016A&A...586A.138P}. In general, B-field lines are believed to provide magnetic support by restricting cross-field gas motion along the filament's length due to the Lorentz force. This facilitates radial collapse of the filaments, as it is easier for the gas to flow along the B-field lines \citep{Mouschovias_1976}. Few observational studies of the filamentary clouds have shown that the B-field dominates the energy budget of the cloud over turbulence and gravity \citep{Chung_2022,Chung_2023}. However, due to the limited resolution of {\it Planck} observations, we are unable to conclusively determine whether gravity or magnetic field dominates in the filament in S284-RE. High resolution polarization observations will be needed to further study the exact role of B-field in the formation and stability of the filament in S284-RE.
\subsection{Clusters of YSOs and their physical properties}
\label{sec:discs1b}
In Section~\ref{sec:datanw}, we identified clusters of YSOs and their associations, and also determined various physical parameters of clusters  YCl1, YCl2, and YCl3. 
The value of $R_{\rm cluster}$ for YCl1 is 1.61 pc, the largest among the three sub-regions, while the aspect ratio, defined as $\frac{R^2_{\rm circ}}{R^2_{\rm cluster}}$ \citep{2014MNRAS.439.3719C}, is 0.31, the lowest value observed. For YCl2, the aspect ratio is 1.89, suggesting a more elongated morphology. Additionally, the mean MST branch length for YCl3, YCl2, and YCl1 are 0.84, 1.17, and 1.79 pc, respectively, indicating that YSOs in YCl1 are more widely separated compared to those in the other clusters.

The $Q$ parameter, defined as the ratio of the normalized mean MST branch length ($\bar{l}_{MST}$) to the normalized mean separation between objects ($\bar{s}$), is essential for evaluating the degree of hierarchical versus radial distributions (see \citealt{2006A&A...449..151S,2014MNRAS.439.3719C,Ascenso2018}). Here, $\bar{l}_{MST}$ is normalized by $\sqrt{A_{\rm cluster}/n_{\rm total}}$, while $\bar{s}$ is normalized by $R_{\rm cluster}$. When $Q>$ 0.8, the objects exhibit a radial distribution, whereas for $Q<$ 0.8, a more fractal distribution is indicated \citep{2004MNRAS.348..589C}.Based on the calculated \( Q \) values listed in Table \ref{tab:mst_parameter}, all three clusters exhibit a fractal distribution, indicating a clumpy structure. Moreover, the farther the \( Q \) value deviates from 0.8, the greater the degree of central concentration. This suggests that clusters with lower \( Q \) values are more irregular and fragmented, while those with higher \( Q \) values tend to be more centrally condensed.

Using the {\it Herschel} $N(\rm H_2)$ map, we also calculated the mass of the area covered by each cluster (see equation~\ref{eq1}), which are listed in Table \ref{tab:mst_parameter}. Among these three clusters, the area covered by the cluster YCl2 has the highest mass, which is 136.35 $M_\odot$.

Jeans fragmentation plays a crucial role in understanding the initial structure of star-forming regions. The Jeans length, $\lambda_J$, which represents the minimum radius necessary for the gravitational collapse of a homogeneous isothermal cloud with mean density $\rho_0$ and temperature T, is given by \citet{2014MNRAS.439.3719C}:
\begin{equation}
    \lambda_{\rm J} = \left(\frac{15 k T}{4 \pi G m_{\rm H} \rho_0}\right)^{1/2}, 
\end{equation}
where $m_{\rm H}$ is the atomic mass of the hydrogen. The mean density $\rho_0$ of a cluster area (gas and stars) with a total mass $M_{\rm T}$ is defined as, 

\begin{equation}
    \rho_0 = \frac{3 M_{\rm T}}{4 \pi R_{\rm H}^3}. 
\end{equation}

The $\lambda_{\rm J}$ values for the clusters range from 0.76 to 4.48 pc (refer to Table \ref{tab:mst_parameter}). By comparing $\lambda_{\rm J}$ with the mean separation between the YSOs, we observed evidence of non-thermally driven fragmentation in YCl1, as the YSO separations are smaller than the Jeans length \citep{2014MNRAS.439.3719C,2020ApJ...891...81P,2024AJ....168...98V}.

Star formation efficiency (SFE) refers to the percentage of gas and dust mass within the $Q_{hull}$ area of a cluster that has been converted into stars \citep[see][]{2008ApJ...688.1142K}. The SFE is calculated in star-forming regions to determine whether there is still molecular gas available that could form new stars. Our results show that the SFE is not correlated with the number of YSOs in each cluster. This absence of correlation is consistent with previous studies \citep{2014MNRAS.439.3719C, 2016AJ....151..126S, 2020ApJ...891...81P}, which propose that feedback processes, such as stellar winds and radiation, are likely to play a significant role primarily in the later stages of cluster evolution, rather than in the early stages of star formation.

Among the clusters, YCl2 has the lowest SFE, suggesting it is the youngest, as indicated by a higher proportion of Class~I YSOs. In contrast, YCl3 appears to be the oldest, as it contains only Class~II YSOs. One may note that according to \citet{evans09}, the average ages of the Class~I and Class~II YSOs are $\sim$0.44 Myr and $\sim$1--3 Myr, respectively. 
\subsection{Clues to ongoing massive star formation and episodic accretion}
\label{sec:discs2}
A combined analysis of data from the {\it Herschel} and {\it Spitzer} facilities has uncovered an embedded filament associated with outflow signatures (see Figure~\ref{fg1}b). 
ALMA 1.38 mm continuum maps reveal continuum sources associated with the {\it Spitzer}-detected outflows, which appear to host the driving sources of the outflows. At least three sub-regions (Rg1, Rg2, and Rg3), where YSOs are distributed, have been investigated. 
 
As mentioned in Section~\ref{sec:data4}, the {\it JWST} data cover two sub-regions (Rg1 and Rg2), and reveal embedded structures in both areas. 
Among the seven H$_{2}$ bipolar outflows (olc1--olc4 and ol1--ol3) traced in the {\it JWST} 470N+F444W image, at least three outflows olc1, ol1, and ol2 stand out as the most distinct and prominent. 
As discussed previously, we find the presence of several H$_{2}$ knots in olc1 and ol1 (see Figures~\ref{fg4} and~\ref{fg5}), and mutlitple bow shocks in H$_{2}$ are found toward olc1 (see Figure~\ref{fg5}). The driving sources of the outflows are also identified (see Section~\ref{sec:data4} for more details and also Figure~\ref{fg5}). In particular,  ALMA continuum source \#2 hosts at least three outflow-driving sources, while ALMA continuum source \#3 hosts two. Both the continuum sources are potential massive star formation candidates, as suggested by the $M$ versus $R_{\rm eff}$ plot/space (see Figure~\ref{fg7}). 
 However, the validity of Kauffmann $\&$ Pillai's criterion for low-metallicity environments remains uncertain. As a result, it is unclear whether all the sources that meet this criterion are truly massive star-forming candidates. Interestingly, the driving source of the bipolar outflow olc1 embedded in the ALMA continuum source \#2 (see Figure~\ref{fg7}a and also the red circle in Figure~\ref{fg6}) is a strong candidate for a massive protostar due to the presence of both Br-$\alpha$ (ionized gas) and PAH emission.
 
Recently, \citet{ray23} examined the {\it JWST} high-resolution infrared observations of the HH211 bipolar outflow that is a well-known protostellar jet associated with a Class~0 protostar located in a nearby solar metallicity region. The {\it JWST} data allowed them to uncover the fine details of the outflow's morphology, including its bow shocks and the complex interactions of the outflow with its surrounding environment \citep[see also][]{caratti24}. 
The central protostellar region of the HH211 outflow appears relatively simple. In contrast, the central protostellar region of the outflow olc1 is more complex, marked by Br-$\alpha$ and PAH emissions, suggesting the potential presence of at least one massive star within a cluster environment. 
Outflow activity signals the initiation of the accretion process in forming YSOs \citep{arce07}. 
The detection of multiple knots or bow-shocks within protostellar outflows often points to episodic accretion \citep[e.g.,][]{audard14,frank14,plunkett15,Fedriani_2023,Lee_2024}. 
Rather than accumulating material in a steady flow, protostars gather mass from their disks in bursts. 
Each burst triggers energetic outflows, leading to the formation of knots or shocks that are observed in 
molecular lines or ionized emission. This episodic nature is commonly driven by instabilities in the disk 
or interactions with magnetic fields, which temporarily increase accretion rates \citep[see][for more details]{snell88,bachiller96,shepherd96,zhang05,arce07}. 
The most intriguing structure revealed in the {\it JWST} images is the molecular outflow olc1, characterized by multiple bow shocks and knots. This outflow seems to be driven by a massive star. Recent studies have shown that flaring methanol masers can trace episodic accretion events in young protostars \citep[e.g.,][]{bertout89,hirota18,stecklum21}. Therefore, our findings suggest episodic ejections from the massive star embedded within continuum source \#2, likely driven by accretion events.

\citet{Vorobyov_2020} investigated the early evolution of protostellar disks with varying metallicities (0.01--1.0 $Z_{\odot}$) using numerical hydrodynamics simulations. They found that protostellar accretion rates in low-metallicity environments are highly variable, characterized by intense bursts of episodic accretion driven by inward-migrating gaseous clumps formed by disk fragmentation \citep{2005ApJ...633L.137V, 2015ApJ...805..115V, 2017MNRAS.464L..90M}. In contrast, this behavior is not commonly observed in the protostellar disks of solar-metallicity
%, where accretion tends to be more stable and less episodic 
\citep{spezzi_2012, Demarch_2015}. High resolution JWST images reveal the presence of H$_2$ knots toward S284-RE, which can be explained through episodic accretion.
%, which can be a common signature in low-metallicity environments.} 
%The reduced cooling efficiency in such environments likely slows the accumulation of material around the protostar, resulting in such periodic bursts of accretion \citep{Vorobyov_2020}.}

%
 In low-metallicity environments, the reduced dust content leads to less attenuation of UV photons emitted by massive stars, allowing these photons to travel farther and resulting in a longer mean free path \citep{Israel_2011}. Hence, the UV radiation field becomes diluted and spreads over larger spatial scales.  However, this is not valid in solar-metallicity regions, where dust attenuation is stronger. The extended reach of ionizing photons can create more diffuse and extended H\,{\sc ii} regions. 
%which are less dense than those in high-metallicity environments 
\citep{Inoue_2002, Arthur_2004,Tremblin_2014}. This allows ionized regions to exert pressure on the surrounding neutral gas, compressing it and potentially triggering star formation at greater distances from the ionizing source as compared to high metallicity regions. The reduced destructive feedback and enhanced gas compression suggest that triggered star formation could be more efficient in low-metallicity environments.
% because feedback mechanisms are less destructive and more effective in compressing the sorrounding gas.}

\section{Summary and Conclusions}
\label{sec:conc}
We present a multi-scale, multi-wavelength study of the S284-RE region, associated with an extended S284 H\,{\sc ii} region, a low-metallicity environment. This research uses data sets from the {\it Spitzer}, {\it Herschel}, {\it Planck}, ALMA, and {\it JWST} facilities. 
The key findings of this study are outlined below.
\begin{enumerate}
\item{A filamentary structure (mass $\sim$2402 $M_{\odot}$; length $\sim$8.5 pc) is investigated in S284-RE using the {\it Herschel} column density map, and is classified as a thermally supercritical filament (i.e., $M_\mathrm{l,crit}$ $<$ $M_\mathrm{l,obs}$). The {\it Planck} 850~$\mu$m polarization data hint that the B-field lines appear to be perpendicular to this filament.}
\item{The {\it Spitzer} ratio map of 4.5 $\mu$m/3.6 $\mu$m emission has revealed H$_{2}$ outflow activities toward the filamentary structure in S284-RE. 
The previously identified YSOs from Puga09 are spatially distributed toward this filamentary structure.}
\item{On the basis of the ALMA 1.38 mm continuum emission contours, the distribution of YSOs, and the H$_{2}$ emission traced in the {\it Spitzer} ratio map, three sub-regions (Rg1, Rg2, and Rg3) are identified in the filament structure, where ongoing star formation activity is evident.}
\item{Three clusters of YSOs within the filament are identified and labeled as YCl1, YCl2, and YCl3. 
Cluster YCl1 is associated with the sub-region Rg1, while cluster YCl2 is associated with sub-regions Rg2 and Rg3. 
In particular, YCl1 shows a wider spacing between its YSOs compared to the Jeans length, suggesting the influence of non-thermal fragmentation processes. 
Cluster YCl3 seems to be more evolved, and YCl2 is the youngest.}
\item{The {\it JWST} (F470N+F444W)/F356W ratio map has revealed at least seven H$_{2}$ bipolar outflows, with four (olc1--olc4) detected in Rg1 (or YCl1) 
and three (ol1--ol3) in Rg2 (or YCl2). Several H$_{2}$ knots in olc1 and ol1, and mutlitple bow-shocks in H$_{2}$ are found toward olc1.}
\item{Taking into account the geometry of outflows, the ALMA continuum peaks, and the previously identified YSOs, the driving sources of the outflows in Rg1 and Rg2 are identified.} 
\item{In the central region of the bipolar outflow olc1, a distinct small-scale feature is more apparent in the {\it JWST} F162M, F182M, and F200W images. 
Additionally, the southern direction of this feature stands out in the F356W, F405N, and F470N images, and is associated with the ALMA 1.38 mm continuum emission.}
\item{The $M$-$R_{\rm eff}$ plot for the ALMA continuum sources identified using {\it astrodendro} toward the Rg1, Rg2, and Rg3 sub-regions, indicates that only sources \#2 and \#3 emerge as potential candidates for massive star-formation. The ALMA continuum source \#2 hosts at least three outflow-driving sources, while the continuum source \#3 harbors two outflow-driving sources.}
\item{The driving source of the bipolar outflow olc1, embedded in the ALMA continuum source \#2, is considered a potential candidate for a massive protostar. This is supported by the detection of both Br-$\alpha$ (ionized gas) and PAH emissions, identified through the {\it JWST} (F405N+F444W)/F356W ratio map.}
\item{The existence of multiple knots in the molecular outflows olc1 and ol1 favours the idea of episodic accretion in their outflow-driving sources.}
\end{enumerate} 

Overall, this study reveals a newly identified massive protostar candidate, embedded in a thermally supercritical filament, driving a $\sim$2.7 pc bipolar H$_2$ outflow (olc1) and undergoing episodic accretion.
\section*{Acknowledgments}
We thank the anonymous referee for providing the valuable comments and suggestions, that improved the scientific content of this paper.
The research work at Physical Research Laboratory is funded by the Department of Space, Government of India. 
This paper makes use of the following ALMA data: ADS/JAO.ALMA\#2021.1.01706.S. ALMA is a partnership of ESO (representing its member states), NSF (USA) and NINS (Japan), together with NRC (Canada) , MOST and ASIAA (Taiwan), and KASI (Republic of Korea), in cooperation with the Republic of Chile. The Joint ALMA Observatory is operated by ESO, AUI/NRAO and NAOJ. The JWST data presented in this article were obtained from the Mikulski Archive for Space Telescopes (MAST) at the Space Telescope Science Institute. The specific observations analyzed can be accessed via \dataset[10.17909/ea2v-0g53]{http://dx.doi.org/10.17909/ea2v-0g53}. A.V. acknowledges the financial support of DST-INSPIRE (No.: DST/INSPIRE Fellowship/2019/IF190550). 
%
%%%%%%%%%%Figures

%%%%%%%%%%%%%%%%%%%% REFERENCES %%%%%%%%%%%%%%%%%%

\bibliographystyle{aasjournal}
\bibliography{reference}{}

\begin{thebibliography}{}
\expandafter\ifx\csname natexlab\endcsname\relax\def\natexlab#1{#1}\fi
\providecommand{\url}[1]{\href{#1}{#1}}
\providecommand{\dodoi}[1]{doi:~\href{http://doi.org/#1}{\nolinkurl{#1}}}
\providecommand{\doeprint}[1]{\href{http://ascl.net/#1}{\nolinkurl{http://ascl.net/#1}}}
\providecommand{\doarXiv}[1]{\href{https://arxiv.org/abs/#1}{\nolinkurl{https://arxiv.org/abs/#1}}}

\bibitem[{{Andr{\'e}} {et~al.}(2014{\natexlab{a}}){Andr{\'e}}, {Di Francesco},
  {Ward-Thompson}, {Inutsuka}, {Pudritz}, \& {Pineda}}]{andre14}
{Andr{\'e}}, P., {Di Francesco}, J., {Ward-Thompson}, D., {et~al.}
  2014{\natexlab{a}}, in Protostars and Planets VI, ed. H.~{Beuther}, R.~S.
  {Klessen}, C.~P. {Dullemond}, \& T.~{Henning}, 27,
  \dodoi{10.2458/azu\_uapress\_9780816531240-ch002}

\bibitem[{{Andr{\'e}} {et~al.}(2014{\natexlab{b}}){Andr{\'e}}, {Di Francesco},
  {Ward-Thompson}, {Inutsuka}, {Pudritz}, \& {Pineda}}]{2014prpl.conf...27A}
{Andr{\'e}}, P., {Di Francesco}, J., {Ward-Thompson}, D., {et~al.}
  2014{\natexlab{b}}, in Protostars and Planets VI, ed. H.~{Beuther}, R.~S.
  {Klessen}, C.~P. {Dullemond}, \& T.~{Henning}, 27--51,
  \dodoi{10.2458/azu_uapress_9780816531240-ch002}

\bibitem[{{Andr{\'e}} {et~al.}(2010{\natexlab{a}}){Andr{\'e}}, {Men'shchikov},
  {Bontemps}, {K{\"o}nyves}, {Motte}, {Schneider}, {Didelon}, {Minier},
  {Saraceno}, {Ward-Thompson}, {di Francesco}, {White}, {Molinari}, {Testi},
  {Abergel}, {Griffin}, {Henning}, {Royer}, {Mer{\'\i}n}, {Vavrek}, {Attard},
  {Arzoumanian}, {Wilson}, {Ade}, {Aussel}, {Baluteau}, {Benedettini},
  {Bernard}, {Blommaert}, {Cambr{\'e}sy}, {Cox}, {di Giorgio}, {Hargrave},
  {Hennemann}, {Huang}, {Kirk}, {Krause}, {Launhardt}, {Leeks}, {Le Pennec},
  {Li}, {Martin}, {Maury}, {Olofsson}, {Omont}, {Peretto}, {Pezzuto}, {Prusti},
  {Roussel}, {Russeil}, {Sauvage}, {Sibthorpe}, {Sicilia-Aguilar}, {Spinoglio},
  {Waelkens}, {Woodcraft}, \& {Zavagno}}]{andre10}
{Andr{\'e}}, P., {Men'shchikov}, A., {Bontemps}, S., {et~al.}
  2010{\natexlab{a}}, \aap, 518, L102, \dodoi{10.1051/0004-6361/201014666}

\bibitem[{{Andr{\'e}} {et~al.}(2010{\natexlab{b}}){Andr{\'e}}, {Men'shchikov},
  {Bontemps}, {K{\"o}nyves}, {Motte}, {Schneider}, {Didelon}, {Minier},
  {Saraceno}, {Ward-Thompson}, {di Francesco}, {White}, {Molinari}, {Testi},
  {Abergel}, {Griffin}, {Henning}, {Royer}, {Mer{\'\i}n}, {Vavrek}, {Attard},
  {Arzoumanian}, {Wilson}, {Ade}, {Aussel}, {Baluteau}, {Benedettini},
  {Bernard}, {Blommaert}, {Cambr{\'e}sy}, {Cox}, {di Giorgio}, {Hargrave},
  {Hennemann}, {Huang}, {Kirk}, {Krause}, {Launhardt}, {Leeks}, {Le Pennec},
  {Li}, {Martin}, {Maury}, {Olofsson}, {Omont}, {Peretto}, {Pezzuto}, {Prusti},
  {Roussel}, {Russeil}, {Sauvage}, {Sibthorpe}, {Sicilia-Aguilar}, {Spinoglio},
  {Waelkens}, {Woodcraft}, \& {Zavagno}}]{2010A&A...518L.102A}
---. 2010{\natexlab{b}}, \aap, 518, L102, \dodoi{10.1051/0004-6361/201014666}

\bibitem[{{Arce} {et~al.}(2007){Arce}, {Shepherd}, {Gueth}, {Lee}, {Bachiller},
  {Rosen}, \& {Beuther}}]{arce07}
{Arce}, H.~G., {Shepherd}, D., {Gueth}, F., {et~al.} 2007, in Protostars and
  Planets V, ed. B.~{Reipurth}, D.~{Jewitt}, \& K.~{Keil}, 245,
  \dodoi{10.48550/arXiv.astro-ph/0603071}

\bibitem[{{Arthur} {et~al.}(2004){Arthur}, {Kurtz}, {Franco}, \&
  {Albarr{\'a}n}}]{Arthur_2004}
{Arthur}, S.~J., {Kurtz}, S.~E., {Franco}, J., \& {Albarr{\'a}n}, M.~Y. 2004,
  \apj, 608, 282, \dodoi{10.1086/386366}

\bibitem[{{Arzoumanian} {et~al.}(2013){Arzoumanian}, {Andr{\'e}}, {Peretto}, \&
  {K{\"o}nyves}}]{Arzoumanian_2013}
{Arzoumanian}, D., {Andr{\'e}}, P., {Peretto}, N., \& {K{\"o}nyves}, V. 2013,
  \aap, 553, A119, \dodoi{10.1051/0004-6361/201220822}

\bibitem[{{Ascenso}(2018)}]{Ascenso2018}
{Ascenso}, J. 2018, in Astrophysics and Space Science Library, Vol. 424, The
  Birth of Star Clusters, ed. S.~{Stahler}, 1,
  \dodoi{10.1007/978-3-319-22801-3_1}

\bibitem[{{Audard} {et~al.}(2014){Audard}, {{\'A}brah{\'a}m}, {Dunham},
  {Green}, {Grosso}, {Hamaguchi}, {Kastner}, {K{\'o}sp{\'a}l}, {Lodato},
  {Romanova}, {Skinner}, {Vorobyov}, \& {Zhu}}]{audard14}
{Audard}, M., {{\'A}brah{\'a}m}, P., {Dunham}, M.~M., {et~al.} 2014, in
  Protostars and Planets VI, ed. H.~{Beuther}, R.~S. {Klessen}, C.~P.
  {Dullemond}, \& T.~{Henning}, 387--410,
  \dodoi{10.2458/azu_uapress_9780816531240-ch017}

\bibitem[{{Bachiller}(1996)}]{bachiller96}
{Bachiller}, R. 1996, \araa, 34, 111, \dodoi{10.1146/annurev.astro.34.1.111}

\bibitem[{{Beichman} {et~al.}(2012){Beichman}, {Rieke}, {Eisenstein}, {Greene},
  {Krist}, {McCarthy}, {Meyer}, \& {Stansberry}}]{2012SPIE.8442E..2NB}
{Beichman}, C.~A., {Rieke}, M., {Eisenstein}, D., {et~al.} 2012, in Society of
  Photo-Optical Instrumentation Engineers (SPIE) Conference Series, Vol. 8442,
  Space Telescopes and Instrumentation 2012: Optical, Infrared, and Millimeter
  Wave, ed. M.~C. {Clampin}, G.~G. {Fazio}, H.~A. {MacEwen}, \& J.~{Oschmann},
  Jacobus~M., 84422N, \dodoi{10.1117/12.925447}

\bibitem[{{Bertout}(1989)}]{bertout89}
{Bertout}, C. 1989, \araa, 27, 351, \dodoi{10.1146/annurev.aa.27.090189.002031}

\bibitem[{{Bhadari} {et~al.}(2022){Bhadari}, {Dewangan}, {Ojha}, {Pirogov}, \&
  {Maity}}]{bhadari22}
{Bhadari}, N.~K., {Dewangan}, L.~K., {Ojha}, D.~K., {Pirogov}, L.~E., \&
  {Maity}, A.~K. 2022, \apj, 930, 169, \dodoi{10.3847/1538-4357/ac65e9}

\bibitem[{{Bhadari} {et~al.}(2020){Bhadari}, {Dewangan}, {Pirogov}, \&
  {Ojha}}]{bhadari_2020}
{Bhadari}, N.~K., {Dewangan}, L.~K., {Pirogov}, L.~E., \& {Ojha}, D.~K. 2020,
  \apj, 899, 167, \dodoi{10.3847/1538-4357/aba2c6}

\bibitem[{{Bhadari} {et~al.}(2023){Bhadari}, {Dewangan}, {Pirogov}, {Pazukhin},
  {Zinchenko}, {Maity}, \& {Sharma}}]{bhadari23}
{Bhadari}, N.~K., {Dewangan}, L.~K., {Pirogov}, L.~E., {et~al.} 2023, \mnras,
  526, 4402, \dodoi{10.1093/mnras/stad2981}

\bibitem[{{Billot} {et~al.}(2010){Billot}, {Noriega-Crespo}, {Carey}, {Guieu},
  {Shenoy}, {Paladini}, \& {Latter}}]{billot10}
{Billot}, N., {Noriega-Crespo}, A., {Carey}, S., {et~al.} 2010, \apj, 712, 797,
  \dodoi{10.1088/0004-637X/712/2/797}

\bibitem[{{Caratti o Garatti} {et~al.}(2024){Caratti o Garatti}, {Ray},
  {Kavanagh}, {McCaughrean}, {Gieser}, {Giannini}, {van Dishoeck},
  {Justtanont}, {van Gelder}, {Francis}, {Beuther}, {Tychoniec}, {Nisini},
  {Navarro}, {Devaraj}, {Reyes}, {Nazar}, {Klaassen}, {G{\"u}del}, {Henning},
  {Lagage}, {{\"O}stlin}, {Vandenbussche}, {Waelkens}, \& {Wright}}]{caratti24}
{Caratti o Garatti}, A., {Ray}, T.~P., {Kavanagh}, P.~J., {et~al.} 2024, arXiv
  e-prints, arXiv:2409.16061, \dodoi{10.48550/arXiv.2409.16061}

\bibitem[{{Cartwright} \& {Whitworth}(2004)}]{2004MNRAS.348..589C}
{Cartwright}, A., \& {Whitworth}, A.~P. 2004, \mnras, 348, 589,
  \dodoi{10.1111/j.1365-2966.2004.07360.x}

\bibitem[{{Chavarr{\'\i}a} {et~al.}(2014){Chavarr{\'\i}a}, {Allen}, {Brunt},
  {Hora}, {Muench}, \& {Fazio}}]{2014MNRAS.439.3719C}
{Chavarr{\'\i}a}, L., {Allen}, L., {Brunt}, C., {et~al.} 2014, \mnras, 439,
  3719, \dodoi{10.1093/mnras/stu224}

\bibitem[{{Chavarr{\'\i}a} {et~al.}(2008){Chavarr{\'\i}a}, {Allen}, {Hora},
  {Brunt}, \& {Fazio}}]{chavarria08}
{Chavarr{\'\i}a}, L.~A., {Allen}, L.~E., {Hora}, J.~L., {Brunt}, C.~M., \&
  {Fazio}, G.~G. 2008, \apj, 682, 445, \dodoi{10.1086/588810}

\bibitem[{Chung {et~al.}(2023)Chung, Lee, Kwon, Tafalla, Kim, Soam, \&
  Cho}]{Chung_2023}
Chung, E.~J., Lee, C.~W., Kwon, W., {et~al.} 2023, The Astrophysical Journal,
  951, 68, \dodoi{10.3847/1538-4357/acd540}

\bibitem[{{Chung} {et~al.}(2022){Chung}, {Lee}, {Kwon}, {Yoo}, {Soam}, \&
  {Cho}}]{Chung_2022}
{Chung}, E.~J., {Lee}, C.~W., {Kwon}, W., {et~al.} 2022, \aj, 164, 175,
  \dodoi{10.3847/1538-3881/ac8a43}

\bibitem[{{Churchwell} {et~al.}(2006){Churchwell}, {Povich}, {Allen}, {Taylor},
  {Meade}, {Babler}, {Indebetouw}, {Watson}, {Whitney}, {Wolfire}, {Bania},
  {Benjamin}, {Clemens}, {Cohen}, {Cyganowski}, {Jackson}, {Kobulnicky},
  {Mathis}, {Mercer}, {Stolovy}, {Uzpen}, {Watson}, \& {Wolff}}]{churchwell06}
{Churchwell}, E., {Povich}, M.~S., {Allen}, D., {et~al.} 2006, \apj, 649, 759,
  \dodoi{10.1086/507015}

\bibitem[{{Churchwell} {et~al.}(2007){Churchwell}, {Watson}, {Povich},
  {Taylor}, {Babler}, {Meade}, {Benjamin}, {Indebetouw}, \&
  {Whitney}}]{churchwell07}
{Churchwell}, E., {Watson}, D.~F., {Povich}, M.~S., {et~al.} 2007, \apj, 670,
  428, \dodoi{10.1086/521646}

\bibitem[{Colombo {et~al.}(2018)Colombo, Rosolowsky, Duarte-Cabral, Ginsburg,
  Glenn, Zetterlund, Hernandez, Dempsey, \& Currie}]{Colombo_19}
Colombo, D., Rosolowsky, E., Duarte-Cabral, A., {et~al.} 2018, Monthly Notices
  of the Royal Astronomical Society, 483, 4291, \dodoi{10.1093/mnras/sty3283}

\bibitem[{{Condon} {et~al.}(1991){Condon}, {Broderick}, \&
  {Seielstad}}]{condon91}
{Condon}, J.~J., {Broderick}, J.~J., \& {Seielstad}, G.~A. 1991, \aj, 102,
  2041, \dodoi{10.1086/116026}

\bibitem[{{Cox} {et~al.}(2016){Cox}, {Arzoumanian}, {Andr{\'e}}, {Rygl},
  {Prusti}, {Men'shchikov}, {Royer}, {K{\'o}sp{\'a}l}, {Palmeirim}, {Ribas},
  {K{\"o}nyves}, {Bernard}, {Schneider}, {Bontemps}, {Merin}, {Vavrek}, {Alves
  de Oliveira}, {Didelon}, {Pilbratt}, \& {Waelkens}}]{Ncox_2016}
{Cox}, N.~L.~J., {Arzoumanian}, D., {Andr{\'e}}, P., {et~al.} 2016, \aap, 590,
  A110, \dodoi{10.1051/0004-6361/201527068}

\bibitem[{{Cusano} {et~al.}(2011){Cusano}, {Ripepi}, {Alcal{\'a}}, {Gandolfi},
  {Marconi}, {Degl'Innocenti}, {Palla}, {Guenther}, {Bernabei}, {Covino},
  {Neiner}, {Puga}, \& {Hony}}]{cusano11}
{Cusano}, F., {Ripepi}, V., {Alcal{\'a}}, J.~M., {et~al.} 2011, \mnras, 410,
  227, \dodoi{10.1111/j.1365-2966.2010.17438.x}

\bibitem[{{De Marchi} {et~al.}(2013){De Marchi}, {Beccari}, \&
  {Panagia}}]{Demarch_2015}
{De Marchi}, G., {Beccari}, G., \& {Panagia}, N. 2013, \apj, 775, 68,
  \dodoi{10.1088/0004-637X/775/1/68}

\bibitem[{{Deharveng} {et~al.}(2010){Deharveng}, {Schuller}, {Anderson},
  {Zavagno}, {Wyrowski}, {Menten}, {Bronfman}, {Testi}, {Walmsley}, \&
  {Wienen}}]{deharveng10}
{Deharveng}, L., {Schuller}, F., {Anderson}, L.~D., {et~al.} 2010, \aap, 523,
  A6, \dodoi{10.1051/0004-6361/201014422}

\bibitem[{{Delgado} {et~al.}(2010){Delgado}, {Djupvik}, \&
  {Alfaro}}]{delgado10}
{Delgado}, A.~J., {Djupvik}, A.~A., \& {Alfaro}, E.~J. 2010, \aap, 509, A104,
  \dodoi{10.1051/0004-6361/200912961}

\bibitem[{Dewangan {et~al.}(2017)Dewangan, Baug, Ojha, Janardhan, Devaraj, \&
  Luna}]{Dewangan_2017}
Dewangan, L.~K., Baug, T., Ojha, D.~K., {et~al.} 2017, The Astrophysical
  Journal, 845, 34, \dodoi{10.3847/1538-4357/aa7da2}

\bibitem[{{Dewangan} {et~al.}(2024{\natexlab{a}}){Dewangan}, {Bhadari},
  {Maity}, {Eswaraiah}, {Sharma}, \& {Jadhav}}]{dewangan24}
{Dewangan}, L.~K., {Bhadari}, N.~K., {Maity}, A.~K., {et~al.}
  2024{\natexlab{a}}, \mnras, 527, 5895, \dodoi{10.1093/mnras/stad3384}

\bibitem[{{Dewangan} {et~al.}(2024{\natexlab{b}}){Dewangan}, {Bhadari},
  {Maity}, {Eswaraiah}, {Sharma}, \& {Jadhav}}]{Dewangan_2024}
---. 2024{\natexlab{b}}, \mnras, 527, 5895, \dodoi{10.1093/mnras/stad3384}

\bibitem[{{Dewangan} {et~al.}(2023){Dewangan}, {Bhadari}, {Men'shchikov},
  {Chung}, {Devaraj}, {Lee}, {Maity}, \& {Baug}}]{Dewangan_2023}
{Dewangan}, L.~K., {Bhadari}, N.~K., {Men'shchikov}, A., {et~al.} 2023, \apj,
  946, 22, \dodoi{10.3847/1538-4357/acbccc}

\bibitem[{{Dewangan} {et~al.}(2016){Dewangan}, {Ojha}, {Luna}, {Anandarao},
  {Ninan}, {Mallick}, \& {Mayya}}]{dewangan16}
{Dewangan}, L.~K., {Ojha}, D.~K., {Luna}, A., {et~al.} 2016, \apj, 819, 66,
  \dodoi{10.3847/0004-637X/819/1/66}

\bibitem[{{Dewangan} {et~al.}(2024{\natexlab{c}}){Dewangan}, {Jadhav}, {Maity},
  {Bhadari}, {Sharma}, {Padovani}, {Baug}, {Mayya}, \& {Pandey}}]{dewangan24b}
{Dewangan}, L.~K., {Jadhav}, O.~R., {Maity}, A.~K., {et~al.}
  2024{\natexlab{c}}, \mnras, 528, 3909, \dodoi{10.1093/mnras/stae150}

\bibitem[{{Evans} {et~al.}(2009){Evans}, {Dunham}, {J{\o}rgensen}, {Enoch},
  {Mer{\'\i}n}, {van Dishoeck}, {Alcal{\'a}}, {Myers}, {Stapelfeldt}, {Huard},
  {Allen}, {Harvey}, {van Kempen}, {Blake}, {Koerner}, {Mundy}, {Padgett}, \&
  {Sargent}}]{evans09}
{Evans}, Neal~J., I., {Dunham}, M.~M., {J{\o}rgensen}, J.~K., {et~al.} 2009,
  \apjs, 181, 321, \dodoi{10.1088/0067-0049/181/2/321}

\bibitem[{{Fedriani, R.} {et~al.}(2023){Fedriani, R.}, {Caratti o Garatti, A.},
  {Cesaroni, R.}, {Tan, J. C.}, {Stecklum, B.}, {Moscadelli, L.}, {Koutoulaki,
  M.}, {Cosentino, G.}, \& {Whittle, M.}}]{Fedriani_2023}
{Fedriani, R.}, {Caratti o Garatti, A.}, {Cesaroni, R.}, {et~al.} 2023, aanda,
  676, A107, \dodoi{10.1051/0004-6361/202346736}

\bibitem[{{Frank} {et~al.}(2014){Frank}, {Ray}, {Cabrit}, {Hartigan}, {Arce},
  {Bacciotti}, {Bally}, {Benisty}, {Eisl{\"o}ffel}, {G{\"u}del}, {Lebedev},
  {Nisini}, \& {Raga}}]{frank14}
{Frank}, A., {Ray}, T.~P., {Cabrit}, S., {et~al.} 2014, in Protostars and
  Planets VI, ed. H.~{Beuther}, R.~S. {Klessen}, C.~P. {Dullemond}, \&
  T.~{Henning}, 451--474, \dodoi{10.2458/azu_uapress_9780816531240-ch020}

\bibitem[{{Guieu} {et~al.}(2009){Guieu}, {Rebull}, {Stauffer}, {Hillenbrand},
  {Carpenter}, {Noriega-Crespo}, {Padgett}, {Cole}, {Carey}, {Stapelfeldt}, \&
  {Strom}}]{guieu09}
{Guieu}, S., {Rebull}, L.~M., {Stauffer}, J.~R., {et~al.} 2009, \apj, 697, 787,
  \dodoi{10.1088/0004-637X/697/1/787}

\bibitem[{{Gutermuth} {et~al.}(2009{\natexlab{a}}){Gutermuth}, {Megeath},
  {Myers}, {Allen}, {Pipher}, \& {Fazio}}]{2009ApJS..184...18G}
{Gutermuth}, R.~A., {Megeath}, S.~T., {Myers}, P.~C., {et~al.}
  2009{\natexlab{a}}, \apjs, 184, 18, \dodoi{10.1088/0067-0049/184/1/18}

\bibitem[{{Gutermuth} {et~al.}(2009{\natexlab{b}}){Gutermuth}, {Megeath},
  {Myers}, {Allen}, {Pipher}, \& {Fazio}}]{gutermuth09}
---. 2009{\natexlab{b}}, \apjs, 184, 18, \dodoi{10.1088/0067-0049/184/1/18}

\bibitem[{{Gutermuth} {et~al.}(2008){Gutermuth}, {Myers}, {Megeath}, {Allen},
  {Pipher}, {Muzerolle}, {Porras}, {Winston}, \& {Fazio}}]{gutermuth08}
{Gutermuth}, R.~A., {Myers}, P.~C., {Megeath}, S.~T., {et~al.} 2008, \apj, 674,
  336, \dodoi{10.1086/524722}

\bibitem[{{Habart} {et~al.}(2024){Habart}, {Peeters}, {Bern{\'e}}, {Trahin},
  {Canin}, {Chown}, {Sidhu}, {Van De Putte}, {Alarc{\'o}n}, {Schroetter},
  {Dartois}, {Vicente}, {Abergel}, {Bergin}, {Bernard-Salas}, {Boersma},
  {Bron}, {Cami}, {Cuadrado}, {Dicken}, {Elyajouri}, {Fuente}, {Goicoechea},
  {Gordon}, {Issa}, {Joblin}, {Kannavou}, {Khan}, {Lacinbala}, {Languignon},
  {Le Gal}, {Maragkoudakis}, {Meshaka}, {Okada}, {Onaka}, {Pasquini}, {Pound},
  {Robberto}, {R{\"o}llig}, {Schefter}, {Schirmer}, {Tabone}, {Tielens},
  {Wolfire}, {Zannese}, {Ysard}, {Miville-Deschenes}, {Aleman}, {Allamandola},
  {Auchettl}, {Baratta}, {Bejaoui}, {Bera}, {Black}, {Boulanger}, {Bouwman},
  {Brandl}, {Brechignac}, {Br{\"u}nken}, {Buragohain}, {Burkhardt}, {Candian},
  {Cazaux}, {Cernicharo}, {Chabot}, {Chakraborty}, {Champion}, {Colgan},
  {Cooke}, {Coutens}, {Cox}, {Demyk}, {Meyer}, {Foschino}, {Garc{\'\i}a-Lario},
  {Gavilan}, {Gerin}, {Gottlieb}, {Guillard}, {Gusdorf}, {Hartigan}, {He},
  {Herbst}, {Hornekaer}, {J{\"a}ger}, {Janot-Pacheco}, {Kaufman}, {Kemper},
  {Kendrew}, {Kirsanova}, {Klaassen}, {Kwok}, {Labiano}, {Lai}, {Lee},
  {Lefloch}, {Le Petit}, {Li}, {Linz}, {Mackie}, {Madden}, {Mascetti},
  {McGuire}, {Merino}, {Micelotta}, {Misselt}, {Morse}, {Mulas}, {Neelamkodan},
  {Ohsawa}, {Omont}, {Paladini}, {Palumbo}, {Pathak}, {Pendleton},
  {Petrignani}, {Pino}, {Puga}, {Rangwala}, {Rapacioli}, {Ricca},
  {Roman-Duval}, {Roser}, {Roueff}, {Rouill{\'e}}, {Salama}, {Sales},
  {Sandstrom}, {Sarre}, {Sciamma-O'Brien}, {Sellgren}, {Shenoy}, {Teyssier},
  {Thomas}, {Togi}, {Verstraete}, {Witt}, {Wootten}, {Zettergren}, {Zhang},
  {Zhang}, \& {Zhen}}]{habart24}
{Habart}, E., {Peeters}, E., {Bern{\'e}}, O., {et~al.} 2024, \aap, 685, A73,
  \dodoi{10.1051/0004-6361/202346747}

\bibitem[{{Hildebrand}(1983)}]{hildebrand83}
{Hildebrand}, R.~H. 1983, \qjras, 24, 267

\bibitem[{{Hirota}(2018)}]{hirota18}
{Hirota}, T. 2018, Publication of Korean Astronomical Society, 33, 21,
  \dodoi{10.5303/PKAS.2018.33.2.021}

\bibitem[{{Inoue}(2002)}]{Inoue_2002}
{Inoue}, A.~K. 2002, \apj, 570, 688, \dodoi{10.1086/339788}

\bibitem[{{Israel} \& {Maloney}(2011)}]{Israel_2011}
{Israel}, F.~P., \& {Maloney}, P.~R. 2011, \aap, 531, A19,
  \dodoi{10.1051/0004-6361/201016336}

\bibitem[{{Kalari} \& {Vink}(2015)}]{kalari15}
{Kalari}, V.~M., \& {Vink}, J.~S. 2015, \apj, 800, 113,
  \dodoi{10.1088/0004-637X/800/2/113}

\bibitem[{{Kauffmann} \& {Pillai}(2010)}]{kauffmann10}
{Kauffmann}, J., \& {Pillai}, T. 2010, \apjl, 723, L7,
  \dodoi{10.1088/2041-8205/723/1/L7}

\bibitem[{{Kirk} {et~al.}(2014){Kirk}, {Offner}, \&
  {Redmond}}]{2014MNRAS.439.1765K}
{Kirk}, H., {Offner}, S. S.~R., \& {Redmond}, K.~J. 2014, \mnras, 439, 1765,
  \dodoi{10.1093/mnras/stu052}

\bibitem[{{Koenig} {et~al.}(2008){Koenig}, {Allen}, {Gutermuth}, {Hora},
  {Brunt}, \& {Muzerolle}}]{2008ApJ...688.1142K}
{Koenig}, X.~P., {Allen}, L.~E., {Gutermuth}, R.~A., {et~al.} 2008, \apj, 688,
  1142, \dodoi{10.1086/592322}

\bibitem[{{Kumar Dewangan} \& {Anandarao}(2010)}]{dewangan_2010}
{Kumar Dewangan}, L., \& {Anandarao}, B.~G. 2010, \mnras, 402, 2583,
  \dodoi{10.1111/j.1365-2966.2009.16071.x}

\bibitem[{{Lamers}(2005)}]{larmers_2005}
{Lamers}, H. J.~G.~L.~M. 2005, in IAU Symposium, Vol. 227, Massive Star Birth:
  A Crossroads of Astrophysics, ed. R.~{Cesaroni}, M.~{Felli}, E.~{Churchwell},
  \& M.~{Walmsley}, 303--310, \dodoi{10.1017/S1743921305004679}

\bibitem[{Lee {et~al.}(2024)Lee, Lee, Johnstone, Herczeg, \& Aikawa}]{Lee_2024}
Lee, S., Lee, J.-E., Johnstone, D., Herczeg, G.~J., \& Aikawa, Y. 2024, The
  Astrophysical Journal, 964, 34, \dodoi{10.3847/1538-4357/ad21e3}

\bibitem[{{Maity} {et~al.}(2023){Maity}, {Dewangan}, {Bhadari}, {Ojha}, {Chen},
  \& {Pandey}}]{Maity_23}
{Maity}, A.~K., {Dewangan}, L.~K., {Bhadari}, N.~K., {et~al.} 2023, \mnras,
  523, 5388, \dodoi{10.1093/mnras/stad1644}

\bibitem[{{Men'shchikov}(2021)}]{getsf_2022}
{Men'shchikov}, A. 2021, \aap, 649, A89, \dodoi{10.1051/0004-6361/202039913}

\bibitem[{{Meyer} {et~al.}(2017){Meyer}, {Vorobyov}, {Kuiper}, \&
  {Kley}}]{2017MNRAS.464L..90M}
{Meyer}, D.~M.~A., {Vorobyov}, E.~I., {Kuiper}, R., \& {Kley}, W. 2017, \mnras,
  464, L90, \dodoi{10.1093/mnrasl/slw187}

\bibitem[{{Molinari} {et~al.}(2010){Molinari}, {Swinyard}, {Bally}, {Barlow},
  {Bernard}, {Martin}, {Moore}, {Noriega-Crespo}, {Plume}, {Testi}, {Zavagno},
  {Abergel}, {Ali}, {Anderson}, {Andr{\'e}}, {Baluteau}, {Battersby},
  {Beltr{\'a}n}, {Benedettini}, {Billot}, {Blommaert}, {Bontemps}, {Boulanger},
  {Brand}, {Brunt}, {Burton}, {Calzoletti}, {Carey}, {Caselli}, {Cesaroni},
  {Cernicharo}, {Chakrabarti}, {Chrysostomou}, {Cohen}, {Compiegne}, {de
  Bernardis}, {de Gasperis}, {di Giorgio}, {Elia}, {Faustini}, {Flagey},
  {Fukui}, {Fuller}, {Ganga}, {Garcia-Lario}, {Glenn}, {Goldsmith}, {Griffin},
  {Hoare}, {Huang}, {Ikhenaode}, {Joblin}, {Joncas}, {Juvela}, {Kirk},
  {Lagache}, {Li}, {Lim}, {Lord}, {Marengo}, {Marshall}, {Masi}, {Massi},
  {Matsuura}, {Minier}, {Miville-Desch{\^e}nes}, {Montier}, {Morgan}, {Motte},
  {Mottram}, {M{\"u}ller}, {Natoli}, {Neves}, {Olmi}, {Paladini}, {Paradis},
  {Parsons}, {Peretto}, {Pestalozzi}, {Pezzuto}, {Piacentini}, {Piazzo},
  {Polychroni}, {Pomar{\`e}s}, {Popescu}, {Reach}, {Ristorcelli}, {Robitaille},
  {Robitaille}, {Rod{\'o}n}, {Roy}, {Royer}, {Russeil}, {Saraceno}, {Sauvage},
  {Schilke}, {Schisano}, {Schneider}, {Schuller}, {Schulz}, {Sibthorpe},
  {Smith}, {Smith}, {Spinoglio}, {Stamatellos}, {Strafella}, {Stringfellow},
  {Sturm}, {Taylor}, {Thompson}, {Traficante}, {Tuffs}, {Umana}, {Valenziano},
  {Vavrek}, {Veneziani}, {Viti}, {Waelkens}, {Ward-Thompson}, {White},
  {Wilcock}, {Wyrowski}, {Yorke}, \& {Zhang}}]{molinari10}
{Molinari}, S., {Swinyard}, B., {Bally}, J., {et~al.} 2010, \aap, 518, L100,
  \dodoi{10.1051/0004-6361/201014659}

\bibitem[{{Motte} {et~al.}(2018){Motte}, {Bontemps}, \& {Louvet}}]{Motte+2018}
{Motte}, F., {Bontemps}, S., \& {Louvet}, F. 2018, \araa, 56, 41,
  \dodoi{10.1146/annurev-astro-091916-055235}

\bibitem[{{Mouschovias}(1976)}]{Mouschovias_1976}
{Mouschovias}, T.~C. 1976, \apj, 206, 753, \dodoi{10.1086/154436}

\bibitem[{{Palmeirim, P.} {et~al.}(2013){Palmeirim, P.}, {André, Ph.}, {Kirk,
  J.}, {Ward-Thompson, D.}, {Arzoumanian, D.}, {Könyves, V.}, {Didelon, P.},
  {Schneider, N.}, {Benedettini, M.}, {Bontemps, S.}, {Di Francesco, J.},
  {Elia, D.}, {Griffin, M.}, {Hennemann, M.}, {Hill, T.}, {Martin, P. G.},
  {Men’shchikov, A.}, {Molinari, S.}, {Motte, F.}, {Nguyen Luong, Q.},
  {Nutter, D.}, {Peretto, N.}, {Pezzuto, S.}, {Roy, A.}, {Rygl, K. L. J.},
  {Spinoglio, L.}, \& {White, G. L.}}]{Palmeirim_2013}
{Palmeirim, P.}, {André, Ph.}, {Kirk, J.}, {et~al.} 2013, AandA, 550, A38,
  \dodoi{10.1051/0004-6361/201220500}

\bibitem[{{Pandey} {et~al.}(2020){Pandey}, {Sharma}, {Panwar}, {Dewangan},
  {Ojha}, {Bisen}, {Sinha}, {Ghosh}, \& {Pandey}}]{2020ApJ...891...81P}
{Pandey}, R., {Sharma}, S., {Panwar}, N., {et~al.} 2020, \apj, 891, 81,
  \dodoi{10.3847/1538-4357/ab6dc7}

\bibitem[{{Planck Collaboration} {et~al.}(2016{\natexlab{a}}){Planck
  Collaboration}, {Adam}, {Ade}, {Aghanim}, {Akrami}, {Alves}, {Arg{\"u}eso},
  {Arnaud}, {Arroja}, {Ashdown}, {Aumont}, {Baccigalupi}, {Ballardini},
  {Banday}, {Barreiro}, {Bartlett}, {Bartolo}, {Basak}, {Battaglia},
  {Battaner}, {Battye}, {Benabed}, {Beno{\^\i}t}, {Benoit-L{\'e}vy}, {Bernard},
  {Bersanelli}, {Bertincourt}, {Bielewicz}, {Bikmaev}, {Bock}, {B{\"o}hringer},
  {Bonaldi}, {Bonavera}, {Bond}, {Borrill}, {Bouchet}, {Boulanger}, {Bucher},
  {Burenin}, {Burigana}, {Butler}, {Calabrese}, {Cardoso}, {Carvalho},
  {Casaponsa}, {Castex}, {Catalano}, {Challinor}, {Chamballu}, {Chary},
  {Chiang}, {Chluba}, {Chon}, {Christensen}, {Church}, {Clemens}, {Clements},
  {Colombi}, {Colombo}, {Combet}, {Comis}, {Contreras}, {Couchot}, {Coulais},
  {Crill}, {Cruz}, {Curto}, {Cuttaia}, {Danese}, {Davies}, {Davis}, {de
  Bernardis}, {de Rosa}, {de Zotti}, {Delabrouille}, {Delouis}, {D{\'e}sert},
  {Di Valentino}, {Dickinson}, {Diego}, {Dolag}, {Dole}, {Donzelli},
  {Dor{\'e}}, {Douspis}, {Ducout}, {Dunkley}, {Dupac}, {Efstathiou},
  {Eisenhardt}, {Elsner}, {En{\ss}lin}, {Eriksen}, {Falgarone}, {Fantaye},
  {Farhang}, {Feeney}, {Fergusson}, {Fernandez-Cobos}, {Feroz}, {Finelli},
  {Florido}, {Forni}, {Frailis}, {Fraisse}, {Franceschet}, {Franceschi},
  {Frejsel}, {Frolov}, {Galeotta}, {Galli}, {Ganga}, {Gauthier},
  {G{\'e}nova-Santos}, {Gerbino}, {Ghosh}, {Giard}, {Giraud-H{\'e}raud},
  {Giusarma}, {Gjerl{\o}w}, {Gonz{\'a}lez-Nuevo}, {G{\'o}rski}, {Grainge},
  {Gratton}, {Gregorio}, {Gruppuso}, {Gudmundsson}, {Hamann}, {Handley},
  {Hansen}, {Hanson}, {Harrison}, {Heavens}, {Helou}, {Henrot-Versill{\'e}},
  {Hern{\'a}ndez-Monteagudo}, {Herranz}, {Hildebrandt}, {Hivon}, {Hobson},
  {Holmes}, {Hornstrup}, {Hovest}, {Huang}, {Huffenberger}, {Hurier},
  {Ili{\'c}}, {Jaffe}, {Jaffe}, {Jin}, {Jones}, {Juvela}, {Karakci},
  {Keih{\"a}nen}, {Keskitalo}, {Khamitov}, {Kiiveri}, {Kim}, {Kisner},
  {Kneissl}, {Knoche}, {Knox}, {Krachmalnicoff}, {Kunz}, {Kurki-Suonio},
  {Lacasa}, {Lagache}, {L{\"a}hteenm{\"a}ki}, {Lamarre}, {Langer}, {Lasenby},
  {Lattanzi}, {Lawrence}, {Le Jeune}, {Leahy}, {Lellouch}, {Leonardi},
  {Le{\'o}n-Tavares}, {Lesgourgues}, {Levrier}, {Lewis}, {Liguori}, {Lilje},
  {Lilley}, {Linden-V{\o}rnle}, {Lindholm}, {Liu}, {L{\'o}pez-Caniego},
  {Lubin}, {Ma}, {Mac{\'\i}as-P{\'e}rez}, {Maggio}, {Maino}, {Mak},
  {Mandolesi}, {Mangilli}, {Marchini}, {Marcos-Caballero}, {Marinucci},
  {Maris}, {Marshall}, {Martin}, {Martinelli}, {Mart{\'\i}nez-Gonz{\'a}lez},
  {Masi}, {Matarrese}, {Mazzotta}, {McEwen}, {McGehee}, {Mei}, {Meinhold},
  {Melchiorri}, {Melin}, {Mendes}, {Mennella}, {Migliaccio}, {Mikkelsen},
  {Millea}, {Mitra}, {Miville-Desch{\^e}nes}, {Molinari}, {Moneti}, {Montier},
  {Moreno}, {Morgante}, {Mortlock}, {Moss}, {Mottet}, {M{\"u}nchmeyer},
  {Munshi}, {Murphy}, {Narimani}, {Naselsky}, {Nastasi}, {Nati}, {Natoli},
  {Negrello}, {Netterfield}, {N{\o}rgaard-Nielsen}, {Noviello}, {Novikov},
  {Novikov}, {Olamaie}, {Oppermann}, {Orlando}, {Oxborrow}, {Paci}, {Pagano},
  {Pajot}, {Paladini}, {Pandolfi}, {Paoletti}, {Partridge}, {Pasian},
  {Patanchon}, {Pearson}, {Peel}, {Peiris}, {Pelkonen}, {Perdereau}, {Perotto},
  {Perrott}, {Perrotta}, {Pettorino}, {Piacentini}, {Piat}, {Pierpaoli},
  {Pietrobon}, {Plaszczynski}, {Pogosyan}, {Pointecouteau}, {Polenta}, {Popa},
  {Pratt}, {Pr{\'e}zeau}, {Prunet}, {Puget}, {Rachen}, {Racine}, {Reach},
  {Rebolo}, {Reinecke}, {Remazeilles}, {Renault}, {Renzi}, {Ristorcelli},
  {Rocha}, {Roman}, {Romelli}, {Rosset}, {Rossetti}, {Rotti}, {Roudier},
  {Rouill{\'e} d'Orfeuil}, {Rowan-Robinson}, {Rubi{\~n}o-Mart{\'\i}n},
  {Ruiz-Granados}, {Rumsey}, {Rusholme}, {Said}, {Salvatelli}, {Salvati},
  {Sandri}, {Sanghera}, {Santos}, {Saunders}, {Sauv{\'e}}, {Savelainen},
  {Savini}, {Schaefer}, {Schammel}, {Scott}, {Seiffert}, {Serra}, {Shellard},
  {Shimwell}, {Shiraishi}, {Smith}, {Souradeep}, {Spencer}, {Spinelli},
  {Stanford}, {Stern}, {Stolyarov}, {Stompor}, {Strong}, {Sudiwala}, {Sunyaev},
  {Sutter}, {Sutton}, {Suur-Uski}, {Sygnet}, {Tauber}, {Tavagnacco}, {Terenzi},
  {Texier}, {Toffolatti}, {Tomasi}, {Tornikoski}, {Tramonte}, {Tristram},
  {Troja}, {Trombetti}, {Tucci}, {Tuovinen}, {T{\"u}rler}, {Umana},
  {Valenziano}, {Valiviita}, {Van Tent}, {Vassallo}, {Vibert}, {Vidal}, {Viel},
  {Vielva}, {Villa}, {Wade}, {Walter}, {Wandelt}, {Watson}, {Wehus},
  {Welikala}, {Weller}, {White}, {White}, {Wilkinson}, {Yvon}, {Zacchei},
  {Zibin}, \& {Zonca}}]{planck16}
{Planck Collaboration}, {Adam}, R., {Ade}, P.~A.~R., {et~al.}
  2016{\natexlab{a}}, \aap, 594, A1, \dodoi{10.1051/0004-6361/201527101}

\bibitem[{{Planck Collaboration} {et~al.}(2016{\natexlab{b}}){Planck
  Collaboration}, {Ade}, {Aghanim}, {Alves}, {Arnaud}, {Arzoumanian},
  {Ashdown}, {Aumont}, {Baccigalupi}, {Banday}, {Barreiro}, {Bartolo},
  {Battaner}, {Benabed}, {Beno{\^\i}t}, {Benoit-L{\'e}vy}, {Bernard},
  {Bersanelli}, {Bielewicz}, {Bock}, {Bonavera}, {Bond}, {Borrill}, {Bouchet},
  {Boulanger}, {Bracco}, {Burigana}, {Calabrese}, {Cardoso}, {Catalano},
  {Chiang}, {Christensen}, {Colombo}, {Combet}, {Couchot}, {Crill}, {Curto},
  {Cuttaia}, {Danese}, {Davies}, {Davis}, {de Bernardis}, {de Rosa}, {de
  Zotti}, {Delabrouille}, {Dickinson}, {Diego}, {Dole}, {Donzelli}, {Dor{\'e}},
  {Douspis}, {Ducout}, {Dupac}, {Efstathiou}, {Elsner}, {En{\ss}lin},
  {Eriksen}, {Falceta-Gon{\c{c}}alves}, {Falgarone}, {Ferri{\`e}re}, {Finelli},
  {Forni}, {Frailis}, {Fraisse}, {Franceschi}, {Frejsel}, {Galeotta}, {Galli},
  {Ganga}, {Ghosh}, {Giard}, {Gjerl{\o}w}, {Gonz{\'a}lez-Nuevo}, {G{\'o}rski},
  {Gregorio}, {Gruppuso}, {Gudmundsson}, {Guillet}, {Harrison}, {Helou},
  {Hennebelle}, {Henrot-Versill{\'e}}, {Hern{\'a}ndez-Monteagudo}, {Herranz},
  {Hildebrandt}, {Hivon}, {Holmes}, {Hornstrup}, {Huffenberger}, {Hurier},
  {Jaffe}, {Jaffe}, {Jones}, {Juvela}, {Keih{\"a}nen}, {Keskitalo}, {Kisner},
  {Knoche}, {Kunz}, {Kurki-Suonio}, {Lagache}, {Lamarre}, {Lasenby},
  {Lattanzi}, {Lawrence}, {Leonardi}, {Levrier}, {Liguori}, {Lilje},
  {Linden-V{\o}rnle}, {L{\'o}pez-Caniego}, {Lubin}, {Mac{\'\i}as-P{\'e}rez},
  {Maino}, {Mandolesi}, {Mangilli}, {Maris}, {Martin},
  {Mart{\'\i}nez-Gonz{\'a}lez}, {Masi}, {Matarrese}, {Melchiorri}, {Mendes},
  {Mennella}, {Migliaccio}, {Miville-Desch{\^e}nes}, {Moneti}, {Montier},
  {Morgante}, {Mortlock}, {Munshi}, {Murphy}, {Naselsky}, {Nati},
  {Netterfield}, {Noviello}, {Novikov}, {Novikov}, {Oppermann}, {Oxborrow},
  {Pagano}, {Pajot}, {Paladini}, {Paoletti}, {Pasian}, {Perotto}, {Pettorino},
  {Piacentini}, {Piat}, {Pierpaoli}, {Pietrobon}, {Plaszczynski},
  {Pointecouteau}, {Polenta}, {Ponthieu}, {Pratt}, {Prunet}, {Puget}, {Rachen},
  {Reinecke}, {Remazeilles}, {Renault}, {Renzi}, {Ristorcelli}, {Rocha},
  {Rossetti}, {Roudier}, {Rubi{\~n}o-Mart{\'\i}n}, {Rusholme}, {Sandri},
  {Santos}, {Savelainen}, {Savini}, {Scott}, {Soler}, {Stolyarov}, {Sudiwala},
  {Sutton}, {Suur-Uski}, {Sygnet}, {Tauber}, {Terenzi}, {Toffolatti}, {Tomasi},
  {Tristram}, {Tucci}, {Umana}, {Valenziano}, {Valiviita}, {Van Tent},
  {Vielva}, {Villa}, {Wade}, {Wandelt}, {Wehus}, {Ysard}, {Yvon}, \&
  {Zonca}}]{2016A&A...586A.138P}
{Planck Collaboration}, {Ade}, P.~A.~R., {Aghanim}, N., {et~al.}
  2016{\natexlab{b}}, aap, 586, A138, \dodoi{10.1051/0004-6361/201525896}

\bibitem[{{Plunkett} {et~al.}(2015){Plunkett}, {Arce}, {Mardones}, {van
  Dokkum}, {Dunham}, {Fern{\'a}ndez-L{\'o}pez}, {Gallardo}, \&
  {Corder}}]{plunkett15}
{Plunkett}, A.~L., {Arce}, H.~G., {Mardones}, D., {et~al.} 2015, \nat, 527, 70,
  \dodoi{10.1038/nature15702}

\bibitem[{{Puga} {et~al.}(2009){Puga}, {Hony}, {Neiner}, {Lenorzer}, {Hubert},
  {Waters}, {Cusano}, \& {Ripepi}}]{puga09}
{Puga}, E., {Hony}, S., {Neiner}, C., {et~al.} 2009, \aap, 503, 107,
  \dodoi{10.1051/0004-6361/200810664}

\bibitem[{{Ray} {et~al.}(2023){Ray}, {McCaughrean}, {Caratti o Garatti},
  {Kavanagh}, {Justtanont}, {van Dishoeck}, {Reitsma}, {Beuther}, {Francis},
  {Gieser}, {Klaassen}, {Perotti}, {Tychoniec}, {van Gelder}, {Colina},
  {Greve}, {G{\"u}del}, {Henning}, {Lagage}, {{\"O}stlin}, {Vandenbussche},
  {Waelkens}, \& {Wright}}]{ray23}
{Ray}, T.~P., {McCaughrean}, M.~J., {Caratti o Garatti}, A., {et~al.} 2023,
  \nat, 622, 48, \dodoi{10.1038/s41-023-06551-1}

\bibitem[{{Rieke} {et~al.}(2005){Rieke}, {Kelly}, \&
  {Horner}}]{2005SPIE.5904....1R}
{Rieke}, M.~J., {Kelly}, D., \& {Horner}, S. 2005, in Society of Photo-Optical
  Instrumentation Engineers (SPIE) Conference Series, Vol. 5904, Cryogenic
  Optical Systems and Instruments XI, ed. J.~B. {Heaney} \& L.~G. {Burriesci},
  1--8, \dodoi{10.1117/12.615554}

\bibitem[{{Rigby} {et~al.}(2019){Rigby}, {Moore}, {Eden}, {Urquhart}, {Ragan},
  {Peretto}, {Plume}, {Thompson}, {Currie}, \& {Park}}]{Rigby_19}
{Rigby}, A.~J., {Moore}, T.~J.~T., {Eden}, D.~J., {et~al.} 2019, \aap, 632,
  A58, \dodoi{10.1051/0004-6361/201935236}

\bibitem[{{Rosen} {et~al.}(2020){Rosen}, {Offner}, {Sadavoy}, {Bhandare},
  {V{\'a}zquez-Semadeni}, \& {Ginsburg}}]{rosen20}
{Rosen}, A.~L., {Offner}, S. S.~R., {Sadavoy}, S.~I., {et~al.} 2020, \ssr, 216,
  62, \dodoi{10.1007/s11214-020-00688-5}

\bibitem[{{Rosolowsky} {et~al.}(2008){Rosolowsky}, {Pineda}, {Kauffmann}, \&
  {Goodman}}]{Rosolowsky08}
{Rosolowsky}, E.~W., {Pineda}, J.~E., {Kauffmann}, J., \& {Goodman}, A.~A.
  2008, \apj, 679, 1338, \dodoi{10.1086/587685}

\bibitem[{{Rémy-Ruyer} {et~al.}(2014){Rémy-Ruyer}, {Madden}, {Galliano},
  {Galametz}, {Takeuchi}, {Asano}, {Zhukovska}, {Lebouteiller}, {Cormier},
  {Jones}, {Bocchio}, {Baes}, {Bendo}, {Boquien}, {Boselli}, {DeLooze},
  {Doublier-Pritchard}, {Hughes}, {Karczewski}, \& {Spinoglio}}]{Remy_2014}
{Rémy-Ruyer}, A., {Madden}, S.~C., {Galliano}, F., {et~al.} 2014, \aap, 563,
  A31, \dodoi{10.1051/0004-6361/201322803}

\bibitem[{Sanhueza {et~al.}(2017)Sanhueza, Jackson, Zhang, Guzmán, Lu,
  Stephens, Wang, \& Tatematsu}]{Sanhueza_2017}
Sanhueza, P., Jackson, J.~M., Zhang, Q., {et~al.} 2017, The Astrophysical
  Journal, 841, 97, \dodoi{10.3847/1538-4357/aa6ff8}

\bibitem[{{Schmeja} \& {Klessen}(2006)}]{2006A&A...449..151S}
{Schmeja}, S., \& {Klessen}, R.~S. 2006, \aap, 449, 151,
  \dodoi{10.1051/0004-6361:20054464}

\bibitem[{{Sharma} {et~al.}(2016){Sharma}, {Pandey}, {Borissova}, {Ojha},
  {Ivanov}, {Ogura}, {Kobayashi}, {Kurtev}, {Gopinathan}, \& {Kesh
  Yadav}}]{2016AJ....151..126S}
{Sharma}, S., {Pandey}, A.~K., {Borissova}, J., {et~al.} 2016, \aj, 151, 126,
  \dodoi{10.3847/0004-6256/151/5/126}

\bibitem[{{Sharma} {et~al.}(2020){Sharma}, {Ghosh}, {Ojha}, {Pandey}, {Sinha},
  {Pandey}, {Ghosh}, {Panwar}, \& {Pandey}}]{2020MNRAS.498.2309S}
{Sharma}, S., {Ghosh}, A., {Ojha}, D.~K., {et~al.} 2020, \mnras, 498, 2309,
  \dodoi{10.1093/mnras/staa2412}

\bibitem[{{Shepherd} \& {Churchwell}(1996)}]{shepherd96}
{Shepherd}, D.~S., \& {Churchwell}, E. 1996, \apj, 472, 225,
  \dodoi{10.1086/178057}

\bibitem[{{Snell} {et~al.}(1988){Snell}, {Huang}, {Dickman}, \&
  {Claussen}}]{snell88}
{Snell}, R.~L., {Huang}, Y.~L., {Dickman}, R.~L., \& {Claussen}, M.~J. 1988,
  \apj, 325, 853, \dodoi{10.1086/166056}

\bibitem[{{Spezzi} {et~al.}(2012){Spezzi}, {De Marchi}, {Panagia},
  {Sicilia-Aguilar}, \& {Ercolano}}]{spezzi_2012}
{Spezzi}, L., {De Marchi}, G., {Panagia}, N., {Sicilia-Aguilar}, A., \&
  {Ercolano}, B. 2012, \mnras, 421, 78,
  \dodoi{10.1111/j.1365-2966.2011.20130.x}

\bibitem[{{Stecklum} {et~al.}(2021){Stecklum}, {Wolf}, {Linz}, {Caratti o
  Garatti}, {Schmidl}, {Klose}, {Eisl{\"o}ffel}, {Fischer}, {Brogan}, {Burns},
  {Bayandina}, {Cyganowski}, {Gurwell}, {Hunter}, {Hirano}, {Kim}, {MacLeod},
  {Menten}, {Olech}, {Orosz}, {Sobolev}, {Sridharan}, {Surcis}, {Sugiyama},
  {van der Walt}, {Volvach}, \& {Yonekura}}]{stecklum21}
{Stecklum}, B., {Wolf}, V., {Linz}, H., {et~al.} 2021, \aap, 646, A161,
  \dodoi{10.1051/0004-6361/202039645}

\bibitem[{{Szécsi, Dorottya} {et~al.}(2015){Szécsi, Dorottya}, {Langer,
  Norbert}, {Yoon, Sung-Chul}, {Sanyal, Debashis}, {de Mink, Selma}, {Evans,
  Christopher J.}, \& {Dermine, Tyl}}]{szecsi_2015}
{Szécsi, Dorottya}, {Langer, Norbert}, {Yoon, Sung-Chul}, {et~al.} 2015,
  aanda, 581, A15, \dodoi{10.1051/0004-6361/201526617}

\bibitem[{{Tan} {et~al.}(2014){Tan}, {Beltr{\'a}n}, {Caselli}, {Fontani},
  {Fuente}, {Krumholz}, {McKee}, \& {Stolte}}]{tan14}
{Tan}, J.~C., {Beltr{\'a}n}, M.~T., {Caselli}, P., {et~al.} 2014, in Protostars
  and Planets VI, ed. H.~{Beuther}, R.~S. {Klessen}, C.~P. {Dullemond}, \&
  T.~{Henning}, 149, \dodoi{10.2458/azu\_uapress\_9780816531240-ch007}

\bibitem[{{Tremblin} {et~al.}(2014){Tremblin}, {Anderson}, {Didelon}, {Raga},
  {Minier}, {Ntormousi}, {Pettitt}, {Pinto}, {Samal}, {Schneider}, \&
  {Zavagno}}]{Tremblin_2014}
{Tremblin}, P., {Anderson}, L.~D., {Didelon}, P., {et~al.} 2014, \aap, 568, A4,
  \dodoi{10.1051/0004-6361/201423959}

\bibitem[{{Verma} {et~al.}(2023){Verma}, {Sharma}, {Mallick}, {Dewangan},
  {Ojha}, {Yadav}, {Pandey}, {Ghosh}, {Kaur}, {Panwar}, \&
  {Chand}}]{2023ApJ...953..145V}
{Verma}, A., {Sharma}, S., {Mallick}, K.~K., {et~al.} 2023, \apj, 953, 145,
  \dodoi{10.3847/1538-4357/acdeef}

\bibitem[{{Verma} {et~al.}(2024){Verma}, {Sharma}, {Dewangan}, {Ojha},
  {Mallick}, {Yadav}, {Kaur}, {Chand}, {Mamta}, \&
  {Gupta}}]{2024AJ....168...98V}
{Verma}, A., {Sharma}, S., {Dewangan}, L.~K., {et~al.} 2024, \aj, 168, 98,
  \dodoi{10.3847/1538-3881/ad5a8b}

\bibitem[{{Vorobyov} \& {Basu}(2005)}]{2005ApJ...633L.137V}
{Vorobyov}, E.~I., \& {Basu}, S. 2005, \apjl, 633, L137, \dodoi{10.1086/498303}

\bibitem[{{Vorobyov} \& {Basu}(2015)}]{2015ApJ...805..115V}
---. 2015, \apj, 805, 115, \dodoi{10.1088/0004-637X/805/2/115}

\bibitem[{{Vorobyov} {et~al.}(2020){Vorobyov}, {Elbakyan}, {Omukai},
  {Hosokawa}, {Matsukoba}, \& {Guedel}}]{Vorobyov_2020}
{Vorobyov}, E.~I., {Elbakyan}, V.~G., {Omukai}, K., {et~al.} 2020, \aap, 641,
  A72, \dodoi{10.1051/0004-6361/202038354}

\bibitem[{{Whitney} {et~al.}(2011){Whitney}, {Benjamin}, {Meade}, {Babler},
  {Watson}, {Churchwell}, {Robitaille}, {Indebetouw}, \& {GLIMPSE360
  Team}}]{whitney11}
{Whitney}, B., {Benjamin}, R., {Meade}, M., {et~al.} 2011, in American
  Astronomical Society Meeting Abstracts, Vol. 217, American Astronomical
  Society Meeting Abstracts \#217, 241.16

\bibitem[{{Zhang} {et~al.}(2005){Zhang}, {Hunter}, {Brand}, {Sridharan},
  {Cesaroni}, {Molinari}, {Wang}, \& {Kramer}}]{zhang05}
{Zhang}, Q., {Hunter}, T.~R., {Brand}, J., {et~al.} 2005, \apj, 625, 864,
  \dodoi{10.1086/429660}

\bibitem[{{Zinnecker} \& {Yorke}(2007)}]{zinnecker07}
{Zinnecker}, H., \& {Yorke}, H.~W. 2007, \araa, 45, 481,
  \dodoi{10.1146/annurev.astro.44.051905.092549}

\bibitem[{Zucker {et~al.}(2018)Zucker, Chen, \& (co PIs)}]{Zucker_2018}
Zucker, C., Chen, H. H.-H., \& (co PIs). 2018, The Astrophysical Journal, 864,
  152, \dodoi{10.3847/1538-4357/aad3b5}

\end{thebibliography}
%\nocite{*}

%%%%%%%%%%%%%%%%%%%%%%%%%%%%%%%%%%%%%%%%%%%%%%%%%%

\end{document}